\documentclass[11pt,a4paper]{article}
\usepackage{jheppub}
\pdfoutput=1
\usepackage[utf8]{inputenc}
\usepackage{amsfonts,amsmath,amssymb}
\usepackage{graphicx}
\usepackage{multirow}
\usepackage{verbatim}
\usepackage{bm}
\usepackage[textsize=scriptsize,backgroundcolor=red!70,linecolor=red]{todonotes}
\usepackage{paralist}
\usepackage{subfig}
\usepackage[normalem]{ulem}
\usepackage{csquotes}
\usepackage{adjustbox}
\usepackage{orcidlink}

\usepackage{tabularx,ragged2e}

\usepackage{booktabs}  

\newcommand{\ii}{\mathrm{i}}
\newcommand{\ee}{\mathrm{e}}

\newcommand{\chisq}{\ensuremath{\chi^{2}}}

\newcommand{\fform}{\ensuremath{\mathcal{F}}}
\newcommand{\bvec}[1]{\ensuremath{\bm #1}}
\newcommand{\dof}{\ensuremath{\text{d.o.f}}}
\newcommand{\apr}{\ensuremath{\alpha^\prime_P}}

\title{The $U$-Matrix geometrical model for multi-particle production in high-energy hadronic collisions}

\author[a]{Rami Oueslati \orcidlink{0000-0003-2569-4056}}
\author[b]{and Adel Trabelsi \orcidlink{0000-0002-2020-3089}}

\emailAdd{rami.oueslati@uliege.be}
\emailAdd{adel.trabelsi@cern.ch}

\affiliation[a]{Space sciences, Technologies and Astrophysics Research (STAR) Institute, Université de Liège, Bât.~B5a, 4000 Liège,  Belgium}
\affiliation[b]{Tunis El Manar University, Faculty of Sciences of Tunis, Nuclear Physics and High Energy Research Unit, Tunis, Tunisia}
\date{\today}

\abstract{Inspired by the picture portraying the KNO scaling violation as an extension of the geometrical scaling violation, the current study proposes a phenomenological model for multi-particle production in hadron collisions based on the geometrical approach and using the $U$-Matrix unitarization scheme of the scattering amplitude. The model has been fine-tuned and all parameters have been derived from optimal fits to various hadronic multiplicity distributions data in $p+p(\Bar{p})$ collisions across a broad range of energies. The results have revealed that our model furnishes a reasonable description of diverse multiplicity distributions at various energies. Besides, they have demonstrated a pronounced violation of the geometrical scaling, which eventually resulted in a significant violation of the KNO scaling. The study has also analyzed the higher-order moments of the multiplicity distribution. We have observed an unexpected overestimation of the fluctuations and correlations between final state particles with increasing energy, particularly above LHC energy. It is claimed that this overestimation is due to statistical fluctuations embedded in the $U$-matrix scheme. The findings of this study have shed light on the key role of the $U$-matrix scheme in the impact of collision geometry on multi-particle production processes at high energy.}


\begin{document}

\maketitle
\section{\label{sec:intro}Introduction}

Throughout the years, the study of multi-particle production in hadron collisions at high energies has sparked the interest of both theoretical physicists and experimentalists given its significance as it offers valuable insights into the intricate mechanisms underlying the production of particles \cite{VANHOVE1971347, Giacomelli:2009fr, Liu_2022, Koley:2022tqg, DREMIN2001301}. 

Similarly, the hadronic multi-particle production is primordial at ultra-high energies. Indeed, it is necessary to have a solid grasp of it so as to properly interpret air-shower cosmic ray observables, which are obtained through a simulation of a wide range of event generators based on Monte Carlo models. These models have been created and adjusted to describe hadronic multi-particle production in man-made accelerator experiments. Despite some minor differences in their fundamental assumptions, the majority of them employ the eikonal as the scheme of the unitarisation of the scattering amplitude. They must also be internally consistent as we depend on them for extrapolating to ultra-high energy scenarios. Nevertheless, modelling hadronic interactions is not devoid of uncertainties, which gives rise to model-dependent outcomes. Among them, one can cite the mass composition problem and the long-standing muon puzzle, which is regarded as one of the most significant issues in hadronic interaction physics \cite{alves2019open}.

Having said that, advances in the phenomenology and theory of multi-particle production may provide a solution to the aforementioned issues.

The study of multi-particle production may be carried out, particularly, through examining the multiplicity distributions of the produced particles. Their analysis is crucial since the patterns observed in them help reveal the complexity of the collision process, shedding light on the interactions involved in particle creation. In fact, this distribution has been thoroughly investigated by experimental collaborations at the LHC, including ALICE, ATLAS, CMS, and LHCb \cite{Grosse-Oetringhaus_2010}, to enhance our comprehension of the rudimentary characteristics of strong interactions and the behaviour of matter under extreme circumstances as well as to verify theoretical predictions.

Besides, the study of the higher-order moments of the multiplicity distribution is of paramount importance as they provide insight into particle features in high-energy collisions, such as correlation and collective behaviour. Indeed, the behaviour of these moments has been the subject of various experimental studies over a wide range of collision energies \cite{CMS:2019fur, Pandav:2022xxx}, furnishing valuable data which are crucial in constraining theoretical models.  

The multiplicity distribution $P_n(s)$ in hadron collisions refers to the distribution of the number of produced particles in a collision event. It is influenced by various factors, such as the colliding hadrons' energy, the collision geometry, and the underlying dynamics of the interaction. More specifically, it is given by the ratio of the topological cross-section $\sigma_n$ to the inelastic cross-section $\sigma_{in}$. The topological cross-section represents the probability or rate of observing a specific configuration of particles with a particular multiplicity value $n$. It is derived from the scattering amplitudes, which are influenced by the underlying scattering processes and dynamics. The inelastic cross-section, on the other hand, characterizes the probability of observing any final-state interaction or particle production, regardless of the specific configuration or multiplicity. It is closely related to the total cross-section and includes contributions from various interaction channels, including both diffractive and non-diffractive processes.

It is important to note that these cross-sections, and thus the multiplicity distribution $P_n$, cannot yet be calculated within the framework of quantum chromodynamics (QCD). Therefore, our current understanding of multi-particle production dynamics relies primarily on phenomenological approaches as well as certain underlying theoretical principles, which form the basis for a wide range of models ~\cite{Grosse-Oetringhaus_2010}. Needless to say, the continuous enhancement of the theoretical models for particle production is critical so as to maintain consistency and coherence with the foundational principles of the Quantum Field Theory (QFT).

One of the key theoretical tenets in the construction of phenomenological models, such as those based on the geometrical or string approach, is the unitarity constraint \cite{Shih_1997, Beggio:1999, Beggio:2007, BEGGIO_2014}. These models often utilize the eikonal scheme as a means to unitarize the scattering amplitude and describe inclusive multiplicity distributions. While this scheme provides a reasonable description of certain hadronic observables, there are compelling reasons to explore alternative schemes for unitarizing the scattering amplitude, namely the $U$-matrix one.

In a recent study \cite{Bhattacharya:2020lac}, it was shown that despite unexpected agreement in the description of the inelastic $pp$ and $p\bar{p}$ cross-sections with the eikonal and $U$-matrix schemes, a divergence in the individual order-by-order amplitudes can potentially impact the topological cross-section by influencing the relative probabilities of different multiplicity configurations and capturing specific physical processes. 

By the same token, in another study \cite{Vanthieghem:2021akb}, it has been found that the $U$-matrix scheme exhibits a slightly better fit to the single diffractive data at high-energies and a faster growth at ultra-high energies compared to the eikonal scheme. This result implies that the underlying physics and dynamics of the single diffractive scattering are sensitive to the choice of scheme, especially at ultra-high energy, further reinforcing the significance of considering the $U$-matrix. Hence, it is worth noting that if the discrepancy in the single diffractive cross-section propagates to the topological cross-section, it can affect the relative probabilities or rates of observing different multiplicity configurations. Moreover, as the inelastic cross-section encompasses both diffractive and non-diffractive processes, the different behaviours in the single diffractive cross-section between unitarization schemes can contribute to variations in the overall inelastic cross-section. 

It should be noted that the specific impact on the topological and inelastic cross-sections would depend on the detailed correlations and interplay between different interaction channels, including both diffractive and non-diffractive processes. In fact, when hadrons collide at high energies, several interactions and processes take place, leading to the formation of large numbers of particles with various species. Given the complex nature of these interactions, deciphering the contributions from individual channels and identifying the specific processes behind multi-particle production appears to be a daunting task \cite{GayDucati:2007fbs}.

On the whole, the inelastic and topological cross-sections are scheme-dependent, implying that the multiplicity distribution $P_n$ is scheme-dependent as well. This will, therefore, impact the description of the multiplicity distribution in terms of its shape, magnitude, or other characteristic features.

An additional reason for considering an alternative to the eikonal unitarization scheme is related to the geometrical scaling, as supported by experimental observations. This scaling is a regularity established by the ISR measurements of the proton-proton and proton-antiproton scattering and refers to the invariability of the ratio between elastic and total cross-sections. Interestingly, experiments conducted at the CERN (SPS) collider have revealed that this regularity is violated when the energy surpasses the ISR energy range \cite{BOZZO1984392, ALNER1984304}. Furthermore, from a theoretical perspective, it has been shown that the violation of the geometrical scaling is more pronounced with the $U$-matrix scheme than with the eikonal as energy increases \cite{Bhattacharya:2020lac}. It should be emphasized that this remarkable discrepancy has proved to occur mainly when we venture into the extremely high-energy region, roughly near the Grand Unification Scale. In this regard, one of the preliminary objectives of the present study is to examine this disparity within an accessible energy range.

Another regularity put forth by the ISR measurements is the KNO scaling so named after its proponents Koba, Nielsen, and Olesen (KNO) \cite{KOBA1972317}. It refers to the constancy of normalized moments in the distribution of multiplicities. Practically, the KNO function is often employed for the examination of multiplicity distributions in particle collisions. It is denoted as $\Phi(z)$, where $\langle n \rangle$ stands for the average of the multiplicity distribution, and $z=n/\langle n \rangle$ represents the normalized multiplicity. It is noteworthy that the KNO function $\Phi(z)$ tends to be independent of the collision energy $\sqrt{s}$ within the ISR energy range. However, considerable deviations from the KNO scaling start to occur at higher energies, such as those attained at FNAL and LHC.

As a matter of fact, several phenomenological studies have demonstrated the existence of a connection between the geometrical scaling of the profile function and the KNO scaling of the multiplicity distribution \cite{Itabashi, Finkelstein:1988cu, lam:1982}. More precisely, these models have shown that the violation of the geometrical scaling, indicated by an increase in the ratio of elastic to total cross-sections ($\sigma_{el}/\sigma_{tot}$) between ISR and collider energies, is associated with the violation of the KNO scaling across these energy ranges within the eikonal scheme. This raises the question of potential consequences for the violation of the KNO scaling within the context of the $U$-matrix scheme. It is expected that the latter, with its modifications to the scattering amplitude, may introduce novel dynamics and fluctuations that influence the statistical behaviour of particle production. These effects can in turn influence the universal properties assumed in the KNO scaling and significantly lead to a violation or alteration of the constancy of normalized moments.

In view of all that has been mentioned so far, one can assume that considering an alternative adequate unitarization scheme can considerably change our understanding of the description of multi-particle production. In the present study, we propose a phenomenological model for multi-particle production in hadron-hadron collisions that hinges on both the geometrical approach and the picture depicting the KNO scaling violation as an extension of the geometrical scaling violation, using the $U$-Matrix scheme. The chief purposes of the study are to examine the geometrical scaling violation within an accessible energy range, describe the hadronic multiplicity distributions in full-phase space over a wide energy range in $p+p(\Bar{p})$ collisions and to investigate the KNO scaling violation. It also aims at probing the correlation between the final particles and revealing the physics that underlies it. In particular, it seeks to provide valuable insights into the relationship between the violation of both the geometrical and KNO scaling and the mechanism of particle production within the $U$-matrix scheme. 

The remainder of the paper is organized as follows: In Section II, the theoretical background of the multi-particle production model will be outlined. Section III will focus on the explicit model and the data used. Section IV will present and discuss the results. Finally, Section V will summarize the findings and discuss the limitations and implications of this work so as to guide future research.

\section{\label{sec:Theory} The theoretical framework of the Model}

The multiplicity distribution $P_{n}(s)$, i.e., the probability of producing $n$ charged particles in an inelastic $p + p(\bar{p})$ collision at the energy $\sqrt{s}$, is given by

\begin{eqnarray}
P_{n}(s) = \frac{\sigma_{n}(s)}{\sigma_{in}(s)} 
\label{eq:mul_dis}
\end{eqnarray}
where $\sigma_{n}(s)$ and $\sigma_{in}(s)$ are the $n$-particle topological and inelastic cross-sections, respectively, with $\sum_{n} \sigma_{n}(s) = \sigma_{in}(s)$. 

Great efforts have been devoted to explaining why the normalized moments in this multiplicity distribution remain constant in the ISR energy range but diverge from this pattern from the LHC energy range. A possible explanation for this phenomenon might be ascribed to the increasing importance of mini-jets production, resulting from both soft and semi-hard partonic processes, as energy increases. In fact, these mini-jets not only contribute to the rapid growth of hadron-hadron cross-sections, as demonstrated by several geometrical models \cite{FAGUNDES201248, Bhattacharya:2020lac, Vanthieghem:2021akb, Oueslati:2023tjt}, but may also play a crucial role in the violation of the KNO scaling.

It goes without saying that the geometrical models based on the impact parameter space representation provide a good description of various aspects of hadron collisions at high energies. Technically speaking, by considering the impact parameter, which quantifies the distance between the colliding particles' centres, these models actually furnish a solid framework for understanding the initial stages of collisions, allowing a systematic exploration of the collision geometry, ranging from central (small impact parameters) to peripheral (large impact parameters) collisions. Therefore, this geometrical approach links the collision geometry to the underlying physics processes, enabling us to infer certain properties. We particularly assume that this approach can also shed light on the intricate interplay between collision geometry and particle multiplicity distribution.

In addition, as has been mentioned in the previous section, the KNO scaling violation can be perceived as an extension of the geometrical scaling violation to the multi-particle production process. This extension highlights that the dynamics governing particle production become more complex and energy-dependent than what a purely geometrical approach can capture. It also suggests that additional physical processes, such as parton interactions, collective effects, or fluctuations, play a role in shaping the multiplicity distribution as collision energies increase. As a result, a more accurate description of the geometrical scaling violation is needed, which in turn will provide a better description of the violation extension to multi-particle production. In fact, the violation of geometrical scaling in impact parameter space occurs when the initial assumptions about simple geometrical overlap and scaling behaviour break down due to more intricate particle-particle interactions or energy-dependent effects. Consequently, this concept harmoniously aligns with the utilization of the impact parameter space representation, emphasizing the significance of collision geometry. 

Overall, when combined with the geometrical scaling tenets, the geometrical models can be enhanced and thus provide a solid framework for describing the multiplicity distribution and uncovering the underpinning universal features of particle production in high-energy hadron collisions. They also furnish a reliable tool that serves to explain the impact of collision geometry on the multiplicity distribution and to analyze it across a broad range of collision energies and particle species, improving prediction and comparison accuracy in multiple collision systems.

To construct our model highlighting the intrinsic relationship between geometrical scaling and KNO scaling within a geometrical approach, we draw inspiration from \cite{CHOU1982301,  Barshay_PRL, lam:1982, Kuang_1983}, where the essence of the approach lies in expressing the overall hadronic multiplicity distributions in the inelastic channel through a combination of an elementary distribution and the inelastic overlap function. Thus, this approach establishes a direct connection between the fluctuations in multiplicity and the fluctuations in the impact parameter, driven by the intuitive understanding that collisions occurring at shorter distances tend to be more violent and yield a higher number of fragments, thereby leading to increased multiplicities. Therefore, in order to establish a reliable geometrical approach to calculate the multiplicity distribution in the impact parameter space representation, we rely on the unitarity condition of the $S$ matrix. This involves examining the elastic scattering amplitude in the impact parameter $b$ space, expressed by the equation:

\begin{eqnarray}
2 \, \textnormal{Im}[\Gamma (s,b)] = |\Gamma (s,b)|^{2} + G_{in}(s,b) ,
\label{unitcond}
\end{eqnarray}
where $\Gamma (s,b)$ denotes the profile function, i.e., the elastic hadron scattering amplitude, and $G_{in}(s,b)$ represents the inelastic overlap function. By employing the optical theorem, we can obtain 
\begin{equation}
\sigma_{tot}(s)=2\int d^{2}b \; \textnormal{Im}[\Gamma (s,b)], 
\end{equation} and recognizing that the function $G_{in}(s,b)$ signifies the absorption probability associated with each $b$ value, we can derive the total inelastic cross-section 
\begin{equation}
\sigma_{in}(s) = \int d^{2}b \; G_{in}(s,b).    
\end{equation}
Thus, the unitarity condition (\ref{unitcond}) is equivalent to $\sigma_{tot}(s)=\sigma_{el}(s)+\sigma_{in}(s)$, where $\sigma_{el}(s) = \int d^{2}b \, |\Gamma (s,b)|^{2}$. 

Various research papers in the field have already explored this phenomenological procedure based on the impact parameter space representation known as the geometrical or string approach \cite{Beggio:1999, Beggio:2007, BEGGIO_2014, Beggio:2020dry}. In this geometrical approach, wherein incident hadrons are treated as spatially extended objects and their collisions are represented as an ensemble of elementary interactions between quarks and/or gluons, the hadronic multiplicity distribution $P_{n}(s)$ is constructed based on elementary quantities associated with microscopic processes. It follows from this that the overall hadronic multiplicity distribution is estimated through summing contributions emerging from each impact parameter $b$ of the incident hadronic system.

To do this, the topological cross-section $\sigma_{n}$ is decomposed into contributions from each impact parameter $b$ with weight $G_{in}(s, b)$ as follows:
\begin{equation}
\begin{aligned}
\sigma_{n}(s) &\equiv \int d^{2}b \; \sigma_{n}(s,b) \\
& = \int d^{2}b \; G_{in}(s,b) \left[ \frac{\sigma_{n}(s,b)}{\sigma_{in}(s,b)} \right],
\end{aligned}
\end{equation} where $\sigma_{in}(s)\equiv \int d^{2}b \, \sigma_{in}(s,b) = \int d^{2}b \, G_{in}(s,b)$.

The quantity enclosed in brackets $ \sigma_{n}(s,b)/\sigma_{in}(s,b) \equiv p_{n}(s,b)$ can be interpreted as the probability of producing $n$ particles at impact parameter $b$. It accounts for interactions among the elementary components of the colliding hadrons. Keeping in mind that $p_{n}(s,b)$ should scale in KNO sense given its elementary structure, the multiplicity distribution $P_{n}(s)$ (\ref{eq:mul_dis}) can be reformulated as :

\begin{equation}
\begin{aligned}
P_{n}(s) = \frac{1}{\sigma_{in}(s)} \int d^{2}b \, \frac{G_{in}(s,b)}{\langle n(s,b) \rangle }\left[\langle n(s,b) \rangle p_{n}(s,b) \right]
\end{aligned}
\label{eq:mul_dis_1}
\end{equation} where $\langle n(s,b) \rangle$ represents the average number of particles produced at $b$ and $s$.

Let $\Phi(s,z)=\langle n(s) \rangle P_{n}(s)$ be the overall multiplicity distribution in KNO form, where
$z=n(s)/\langle n(s) \rangle $ is the corresponding KNO variable. A multiplicity distribution $\phi(s,\b{z})$, associated with elementary processes occurring at $b$ and $s$, can be written in the form $\phi(s,\b{z}) = \langle n(s,b) \rangle p_{n}(s,b)$, where $\b{z} = n(s)/\langle n(s,b) \rangle$. It is worth noting that both distributions adhere to the standard normalizations \cite{lam:1982} :

\begin{eqnarray}
\int_{0}^{\infty} dz\, \Phi(z) = \int_{0}^{\infty} d\b{z}\, \phi(\b{z}) = 2
\label{normaliz001}
\end{eqnarray}

and

\begin{eqnarray}
\int_{0}^{\infty} dz\, z\, \Phi(z) = \int_{0}^{\infty} d\b{z}\, \b{z}\, \phi(\b{z}) = 2 .
\label{normaliz002}
\end{eqnarray}

In this theoretical framework, the unknown function representing the average number of particles produced at a specific impact parameter $b$ and energy $s$ can be comprehended as the product of two factors: The first factor, \(\langle n(s) \rangle\), representing the general behaviour depicts the average number of particles generated in a collision, regardless of the specific impact parameter. As for the second, it is denoted as  \(f(s,b)\) and perceived as a multiplicity function. It describes the variation of the average number of particles in accordance with the impact parameter \(b\) quantifying how collision geometry affects particle production. The mathematical expression for this factorization is:

\begin{equation}
\langle n(s,b) \rangle = \langle n(s) \rangle f(s,b)
\label{eq:averg_n_bs}
\end{equation} and therefore the expression (\ref{eq:mul_dis_1}) can be rewritten in the KNO form as follows:

\begin{eqnarray}
\Phi (s,z) = \langle n(s) \rangle P_{n}(s) = \frac{1}{\int d^{2}b \, G_{in}(s,b)} \int d^{2}b \, \frac{G_{in}(s,b)}{f(s,b)}\,
\phi \left( \frac{z}{f(s,b)} \right) ,
\label{eq: kno}
\end{eqnarray}

and we also have

\begin{eqnarray}
P_{n}(s) = \frac{1}{\langle n(s) \rangle\int d^{2}b \,  G_{in}(s,b) }
\int d^{2}b \, \frac{G_{in}(s,b)}{f(s,b)}\,
\phi^{(1)} \left( \frac{z}{f(s,b)} \right) ,
\label{eq:master_f}
\end{eqnarray}

The master formula eq.~(\ref{eq:master_f}) highlights the connection between the multiplicity distribution $P_n (s)$ and the unitarisation scheme of the scattering amplitude which can ultimately be established within the inelastic overlap function, emphasizing that this multiplicity is scheme-dependent as stated in the introduction. The full phase space hadronic multiplicity distribution $P_{n}(s)$ is hence constructed by summing contributions from parton–parton collisions occurring at each value of $b$. These interactions give rise to the formation of string-like objects, similar to the string formation described by the Lund model \cite{ARTRU197493}. These strings in turn fragment into hadrons. So, in order to stress the fundamental principle of the string approach, an index labelling the elementary multiplicity distribution $\phi^{(1)}$ has been introduced in the master formula (\ref{eq:master_f}). The essential elements in this formula eq.~(\ref{eq:master_f}), namely the inelastic overlap function, the average number of particles produced at a specific impact parameter $b$ and energy $s$, as well as the elementary multiplicity distribution function, will be explicitly presented in the following section, laying the groundwork for an in-depth analysis of the complex interactions between collision geometry and particle production.

\section{Explicit model and data}

The explicit model presented in this paper for describing High-energy hadronic scattering is theoretically grounded in the geometrical approach and is based on the Reggeon exchanges picture (see, e.g. \cite{Donnachie:2013xia} and references therein). In this picture, hypothetical exchange particles, known as pomerons, mediate interactions between hadrons and the procedure of obtaining the amplitude of a given hadronic process involves summing over all conceivable ways in which pomerons can be exchanged.

Diagrammatically, every pomeron exchange is represented by a line that connects the incoming and outgoing hadrons. Hence, by summing all possible topologies of these diagrams, the total amplitude is determined. Technically, the single-pomeron exchange amplitude, also known as, the Born term can be parameterized as follows:

\begin{equation}
	a(s,t) = g_{p}^2 \, \fform_1(t)^2 \left( \frac{s}{s_0} \right)^{\alpha(t)} \, \xi(t),
	\label{eq:amp_t}
\end{equation} where $g_p$ is the pomeron-proton coupling, $\fform_1(t)$ denotes the proton elastic form factor, and $\xi(t)$ stands for the signature factor that is given by:

\begin{equation}
	\xi(t) = -e^{-i\pi\alpha(t)/2},
	\label{eq:amp_t2}
\end{equation} and the pomeron trajectory, represented by $\alpha(t)$, approximates a straight line:

\begin{equation}
	\alpha(t) = 1 + \epsilon + \apr t.
	\label{eq:amp_t3}
\end{equation}

In the impact-parameter space representation, where the Fourier transform of the amplitude $a(s,t)$ rescaled by $2s$ corresponds to a partial wave, we have:

\begin{equation}
	\chi(s, \mathbf{b}) = \int \frac{\mathrm{d}^2\mathbf{q}}{(2\pi)^2}
	\frac{a(s,t)}{2s} \exp(\mathrm{i} \mathbf{q} \cdot \mathbf{b}).
	\label{eq:Gsb}
\end{equation}

When we venture into high energies, summing amplitudes may lead to a violation of unitarity, especially in the case of multi-pomeron exchanges. Therefore, in order to make the summed amplitudes comply with the unitarity constraint, unitarization techniques come into play by resuming infinite series or introducing additional terms to modify the amplitude's behaviour. As a matter of fact, there is a plethora of techniques in the literature \cite{Selyugin:2007jf,Cudell:2008yb, Drescher:2000ha} whose shared objective is to come up with a consistent approach to summing over various exchange contributions while making sure that the resulting amplitude $\Gamma(b,s)$, i.e., the hadronic profile function, satisfies the unitarity condition and may account for all features of interactions in the context of hadron collisions. Among them, we can cite the eikonal scheme, which is one of the most commonly used methods, positing that the profile function is provided by:
\begin{equation}
	\Gamma_E(s, \bvec b) = \ii \left[ 1 - \ee^{\ii \chi(s, \bvec b)} \right]\,,
	\label{eq:eikx}
\end{equation}

The $U$-matrix scheme is another illustration, which asserts that:
\begin{equation}
\Gamma_U(s, \bvec b) = \frac{\chi(s, \bvec b)}{1 - \ii \,  \chi(s, \bvec b)/2}.
\label{eq:umatx}
\end{equation}

As previously mentioned in the introduction, we will consider that the hadronic profile function $\Gamma(b,s)$ is given by the $U$-Matrix form (\ref{eq:umatx}). This function represents the sum of all $n$-pomeron exchange contributions obtained from the single-pomeron exchange amplitude which is, in turn, related to the expected number $\chi(b,s)$ of interactions between partons of the incident hadrons.

Using both the eikonal and the $U$-matrix schemes as well as a dipole-like form factor for the proton, where $\fform_1(t) = 1/(1-t/t_0)^2$, the parameters $\epsilon$ and $\apr$ describing the pomeron trajectory, the coupling constant $g_p$, and the form-factor scale $t_0$ are adjusted based on a best fit to up-to-date hadron collider data on total, elastic, and inelastic cross-sections. The values of these parameters are provided in Table \ref{tab:el} \cite{Bhattacharya:2020lac}. Following this adequate description of the hadronic profile function, we can determine the inelastic overlap function $G_{in}(s, b)$, needed in our explicit model, by using the equation presented in (\ref{unitcond}).

\begin{table}
\begin{center}
\begin{tabular}{|c||c|c|c|c|c|}
   \hline
   Scheme   & $\epsilon $
            & $\apr $ (GeV$^{-2}$)
            & $ g_{p} $
            & $ t_{0}$ (GeV$^2$)
            & $ \chisq/\dof $\\
            \hline
  U-matrix
            & $ 0.10 \pm 0.01 $
            & $ 0.37 \pm 0.28 $
            & $ 7.5  \pm 0.8 $
            & $ 2.5 \pm 0.6 $
            & $ 1.436 $ \\  
   Eikonal
            & $ 0.11 \pm 0.01 $
            & $ 0.31 \pm 0.19 $
            & $ 7.3  \pm 0.9 $
            & $ 1.9  \pm 0.4$
            & $ 1.442 $ \\
   \hline
\end{tabular}

\end{center}
\caption{\label{tab:el}\chisq/\dof\ and best-fit parameters obtained using the eikonal and U-matrix unitarisation schemes.}
\end{table}

Since the primary objective of this work is to examine the influence of geometrical collisions on multi-particle production processes, we shall propose two hypotheses concerning our choice of the elementary multiplicity distribution. First, it is sufficient for our study to consider that, on average, every string created in parton-parton interactions has the same likelihood of producing a certain number of charged hadrons, which is described by the elementary multiplicity distribution \(\phi^{(1)}(z)\), even though the strings created may have different probabilities of turning into a pair of charged hadrons. Second, despite the fact that the parametrization of the elementary distribution $\phi^{(1)}(z)$ is key to capturing the overall shape of the multiplicity distribution, these two do not necessarily have similar shapes. This is mainly because the overall distribution is obtained by summing contributions from elementary processes at different impact parameters, in the context of this superposition model eq.~(\ref{eq:master_f}). That is to say, the peculiar combination of the individual contributions emerging from distinct impact parameters and which may be having differing shapes results in an overall distribution that reflects these contributions' combined effects and whose characteristics should be represented by their superposition. Hence, we will assume, as a first approximation, that the elementary distribution has the same shape as the overall distribution. More specifically, the choice of the functional form for this elementary distribution is motivated by phenomenological fits to data.

As a matter of fact, the Negative Binomial Distribution (NBD) has proved to provide a good description of the experimental data on the multiplicity distributions in the context of high-energy physics and hadron collisions \cite{Grosse-Oetringhaus_2010}. Therefore, in the present study, the KNO form of the NBD, also known as the Gamma distribution, is adopted for the elementary multiplicity distribution and given by:
\begin{equation}
    \phi^{(1)}(z) = 2 \, \frac{K^K}{\Gamma(K)} \, z^{K-1} \, e^{-Kz},
\end{equation} where \(K\) is a dimensionless parameter. 
 
It should be pointed out that there exists a connection between the average number of particles generated at a specific impact parameter $b$ and energy $s$, $\langle n(s,b)\rangle$, which determines the unknown multiplicity function $f(s,b)$ by the eq.~(\ref{eq:averg_n_bs}), and the Born term in $b$ space $\chi(s, b)$. 

This link can be attributed to the various roles that the Born term plays. To begin with, the multiplicity of generated particles in $b$ space and the Born term are interrelated given that the former can be impacted by the effective interaction of partons within the colliding hadrons and the latter provides a measure of this effective interaction since it represents the overlap of the colliding matter distributions. Not to mention, the Born term depends on the impact parameter. As such, it reflects how strong the interaction between colliding hadrons is at distinct impact parameters. Indeed, the impact-parameter-dependent strength of the interaction between hadrons may have an impact on the multiplicity of produced particles. Secondly, owing that the Born term is a crucial parameter in describing the energy dependence of the scattering amplitude and that particle generation is more likely to occur at higher energies, the Born term is inextricably connected to the average particle production. Thirdly, on the one hand, the imaginary part of the Born term is related to inelastic processes in high-energy collisions and, on the other hand, multi-particle production is often associated with inelastic interactions, so relying on the imaginary part of the Born term will reflect the possibility of inelastic scattering and subsequent particle production. Therefore, the multiplicity function \(f(s,b)\) is constructed to depend on the imaginary part of the Born term, denoted as \(\chi_I(s,b)\). 

Therefore, the connection between the multiplicity function \(f(s,b)\) and the born term is formally defined as follows:

\begin{equation}
f(s,b) = \beta(s)\, [\chi_{I}(s,b)]^{2\lambda},
\label{mul_fuc}
\end{equation} where \(\beta(s)\) is determined by the normalisation condition (\ref{normaliz002}) : 

\begin{equation}
 \beta(s)=\frac
  {\int d^2 b\, G_{in}(s,b)}
  {\int d^2 b\, G_{in}(s,b) \,[\chi_{I}(s,b)]^{2\lambda}}
 \label{eq:beta}
\end{equation} and the exponent \(2\lambda\) in eq.~(\ref{mul_fuc}) introduces a power-law dependence on \(\chi_{I}(s,b)\), suggesting a non-linear relationship between the effective overlap and the particle production in the impact-parameter space.

It should be emphasized that according to the geometrical approach, the phenomenological portrayal of hadronic multiplicity distributions is considerably influenced by the values and behaviours of three key inputs: the inelastic overlap function $G_{in}$, the elementary multiplicity distribution $\phi(z)$, and the $\lambda$ parameter determining the power-law dependence on the effective overlap \(\chi_{I}(s,b)\).
A previous study \cite{Beggio:1999} has shown that changing one of these inputs, while keeping the other two fixed, produces different results across different parameterizations. For instance, if the inputs of the inelastic overlap function are different and $\lambda$, as well as $\phi(z)$, are kept constant, the obtained hadronic multiplicity distributions successfully replicate the experimental data. Interestingly, in all cases, the physical picture is that large multiplicities occur for small impact parameters while peripheral collisions (large b) lead to small multiplicities. It is worth noting that all these results, obtained for different inputs of the inelastic overlap function, are generated using the eikonal scheme. However, in the present study, we employ the $U$-matrix scheme for the reasons mentioned in the previous section. This will eventually allow us to fix our choice of the inelastic overlap function.

In order to present our findings in the following section, it is essential to provide an overview of the model parameters that were determined through data fitting, as well as the specific experimental data employed in our analysis. Since the Born term $\chi_{I}(s,b)$ is completely determined from the best-fit, Table \ref{tab:el}, we see from the master formula for the multiplicity distribution eq. (\ref{eq:master_f}) that the only free parameters are $K$, $\lambda$, and $\langle n(s, b) \rangle$. Further, for our purposes, it is sufficient to fix the value of $K$ and assume $\lambda$ and $\langle n(s, b) \rangle$ as the only fitting parameters. Indeed, in \cite{Beggio:1999}, it was shown that the choice of $K = 10.775$ by assuming a Gamma distribution gives a good description of the charged multiplicity distributions for $e^{+}e^{-}$ annihilation data in a large energy interval. Based on the universality of multiplicities in $e^{+}e^{-}$ and $p + p(\Bar{p})$ collisions, we adopt this choice. 

To determine the values of the parameters $\lambda$ and $\langle n \rangle$, we fix the dimensionless parameter to $K=10.775$ and conduct fits to full phase space $P_{n}$ data in $ p + p(\Bar{p})$ collision across a wide range of energies, specifically at $\sqrt{s} = 30.4, 44.5, 52.6, 62.2, 300, 546, 1000,$ and $1800$ GeV \cite{Breakstone_1984, ALEXOPOULOS1998453}.

With regards to the fitting process, we utilized the Minuit2 class from ROOT \cite{Hatlo_2005} and implemented the MIGRAD algorithm. The primary objective of the fitting procedure was to minimize the $\chi^2$ value and the uncertainties associated with the free parameters were calculated using a $1 \sigma$ confidence level.

\section{\label{sec:result}Results and discussion}

\subsection{Geometrical scaling violation}

\begin{figure}[!htpb]
\centering
\subfloat{\includegraphics[height=0.4\textwidth]{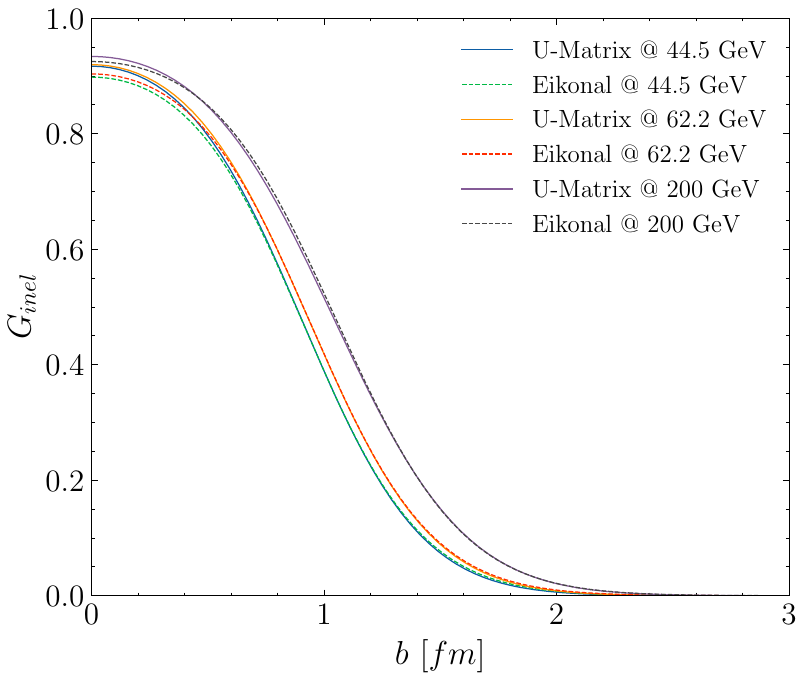}}
\subfloat{\includegraphics[height=0.4\textwidth]{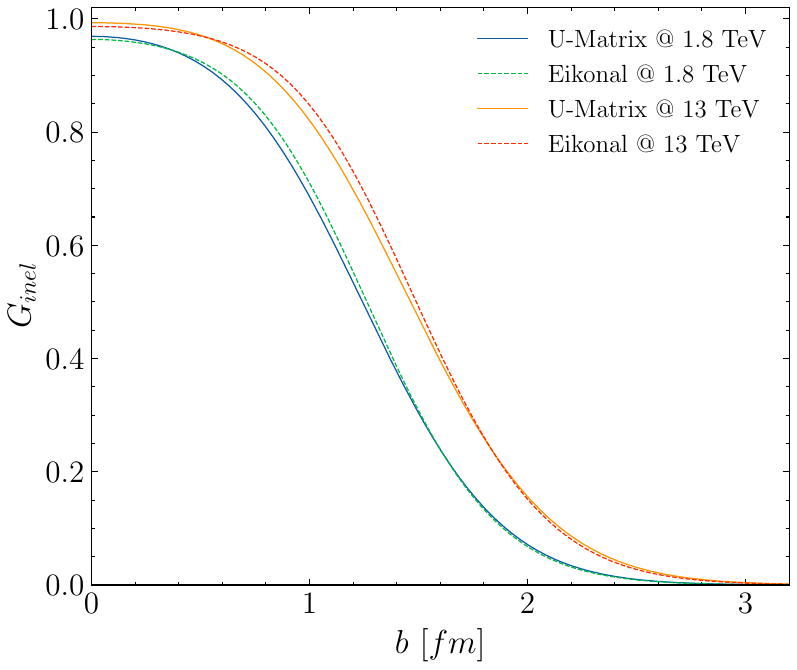}} 
\caption{\label{fig:G_inel} The energy evolution of the inelastic overlap function in the impact parameter space at energies spanning from ISR to LHC levels using both the $U$-matrix and the eikonal schemes.}
\end{figure}

The preliminary objective of this study was to analyze the behaviour of the inelastic overlap function in the impact parameter space as well as the geometrical scaling violation using the $U$-matrix and the eikonal schemes, eqs.~(\ref{eq:eikx}) and (\ref{eq:umatx}), in an attempt to offer valuable insights into the collision geometry and the interaction dynamics of colliding particles. Fig.~\ref{fig:G_inel} depicts the predictions for the energy evolution of the inelastic overlap function, eq.~(\ref{unitcond}), in the impact parameter space at energies spanning from ISR to LHC levels using both schemes.

Looking at Fig.~\ref{fig:G_inel}, it is apparent that this function exhibits a generally comparable pattern in impact parameter space across both schemes, with only minor differences. More specifically, it shows that it is predominantly central, indicating that it has a significant contribution at small impact parameters. This entails that most of the inelastic processes are more probably to take place when the colliding particles pass through one another within close proximity. Moreover, this function tends to decline more slowly with the impact parameter as energy rises. This points out that the inelastic processes become less dependent on the specific spatial distance between colliding particles and have a wider range of impact parameters at which they can happen at higher energies.

\begin{figure}[!htpb]
\centering
\includegraphics[height=0.4\textwidth]{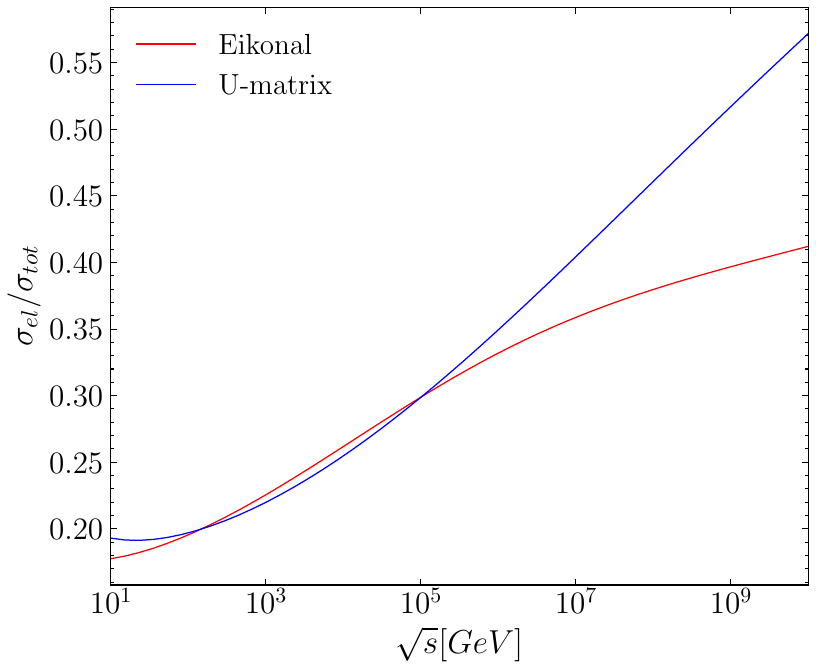}
\caption{\label{fig:ratio} 
Energy evolution of the ratio elastic-to-total cross-section in both cases, the eikonal and $U$-matrix schemes}
\end{figure}

Furthermore, it can be seen from Fig.~\ref{fig:G_inel} that the magnitude of the inelastic overlap function grows with energy at the central impact parameter $b= 0$. This indicates that the higher the energy, the more likely it is that inelastic events will occur at the central impact parameter. Interestingly, as Fig.~\ref{fig:G_inel} shows, there is a significant divergence in the magnitude of the inelastic overlap function obtained from the $U$-matrix and the eikonal schemes. To be more specific, the former yields a greater magnitude than the latter at $b=0$. It is clear from this difference that the chosen scheme has a profound impact on the inelastic processes, particularly on central collisions which will have distinct characteristics or probabilities.

In the same vein, Fig.~\ref{fig:ratio} depicts the energy evolution of the ratio elastic-to-total cross-section using both schemes. As this Figure demonstrates, in both cases there is an overall increase of this ratio as energy rises from $10$ GeV to $10$ TeV, but does so in a non-linear manner, which is indicative of geometrical scaling violation. Interestingly, comparing the behaviour of this ratio in both cases, it is clear that it increases more rapidly in the $U$-matrix case than in the eikonal one, as energy increases, demonstrating a stronger violation of the geometrical scaling and a more intricate behaviour in the former case. 

It is possible to explain the divergence observed in the geometrical scaling violation between the two distinct schemes in terms of the behaviour of the inelastic overlap function in the impact parameter space with respect to energy. To be more precise, this difference lies in the behaviour of the inelastic overlap function at $b = 0$, where we can see from Fig.~\ref{fig:G_inel} that it has a magnitude that grows with energy in both cases. However, the $U$-matrix scheme yields a larger magnitude than the eikonal one. This implies that, in the former case, the inelastic interactions are more prominent, indicating that they have a remarkably and rapidly increasing strength at the central impact parameter with increasing energy, compared to the eikonal case.
 
It can be argued that the discrepancy found in the behavior of the inelastic overlap function as well as in the observed violation of the geometrical scaling between the two schemes can be ascribed to their differing approaches to the unitarization of the scattering amplitude. Indeed, based upon its thorough treatment of processes and its ability to take into account a wider range of interactions and collision dynamics compared to the Eikonal scheme, the U-matrix scheme is likely to be a more advanced mechanism for dealing with scattering processes and capturing the intricate dynamics of hadronic interactions, which is also supported by other results \cite{Oueslati:2023tjt, Vanthieghem:2021akb, Bhattacharya:2020lac}.

On the whole, this result validates the remarkable difference in the geometrical scaling violation between the $U$-matrix and eikonal schemes. Specifically, it supports the claim that this violation is more noticeable with the former scheme when energy levels rise. Notably, this outcome is in agreement with the theoretical prediction provided in \cite{Bhattacharya:2020lac} and further highlights that this discrepancy becomes apparent even before reaching the extremely high-energy region. Furthermore, it reinforces our motivation, as stated in the introductory section, for selecting the $U$-matrix scheme. In fact, the choice of this scheme, in particular, may contribute to answering the question posed in the introduction regarding its potential implications for KNO scaling violation and unravelling the underlying physics behind the multi-particle production mechanism, as will be elaborated in the forthcoming sections.

\subsection{Hadronic Multiplicity Distributions}

\begin{figure}[!htpb]
\centering
\subfloat{\includegraphics[height=0.4\textwidth]{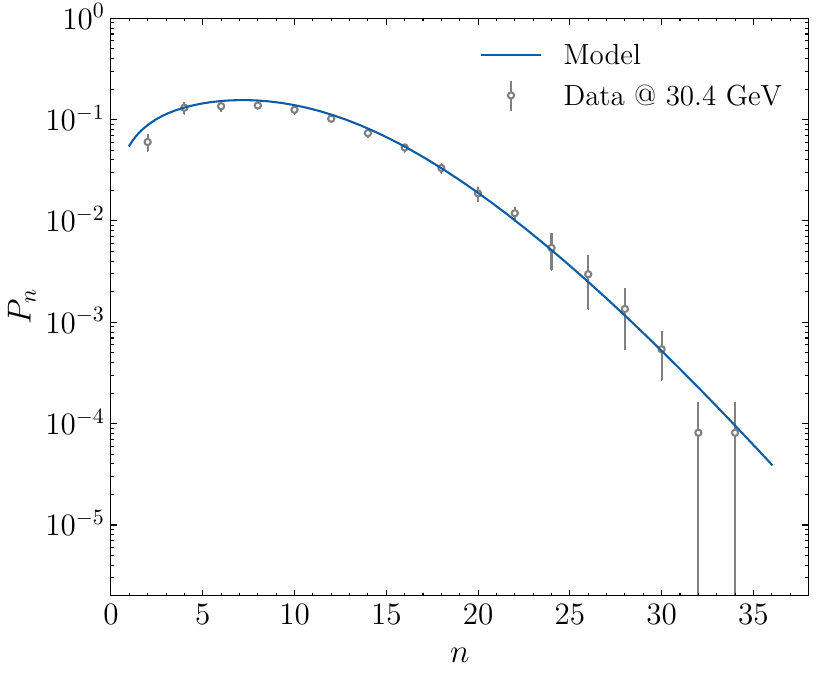}}
\subfloat{\includegraphics[height=0.4\textwidth]{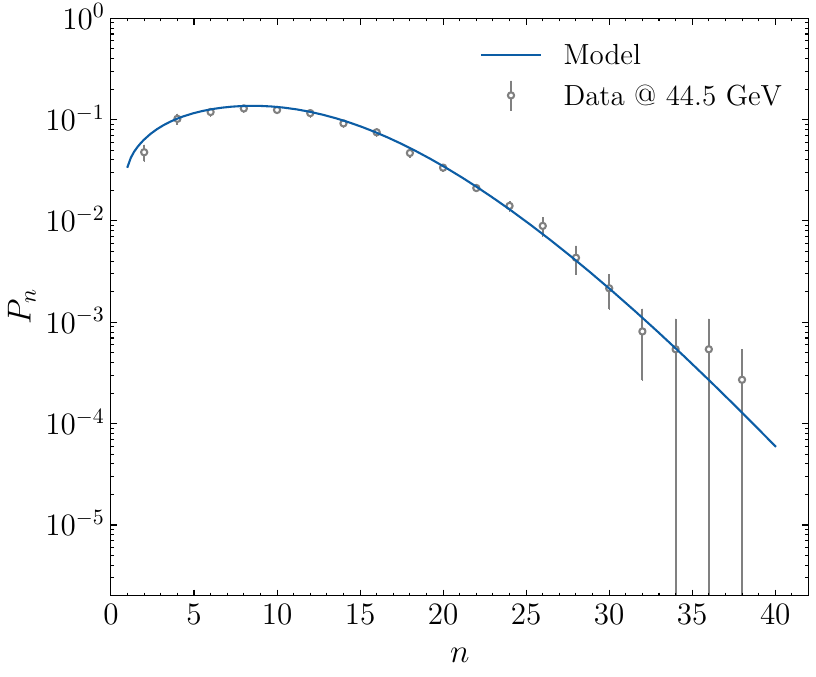}} \\
\subfloat{\includegraphics[height=0.4\textwidth]{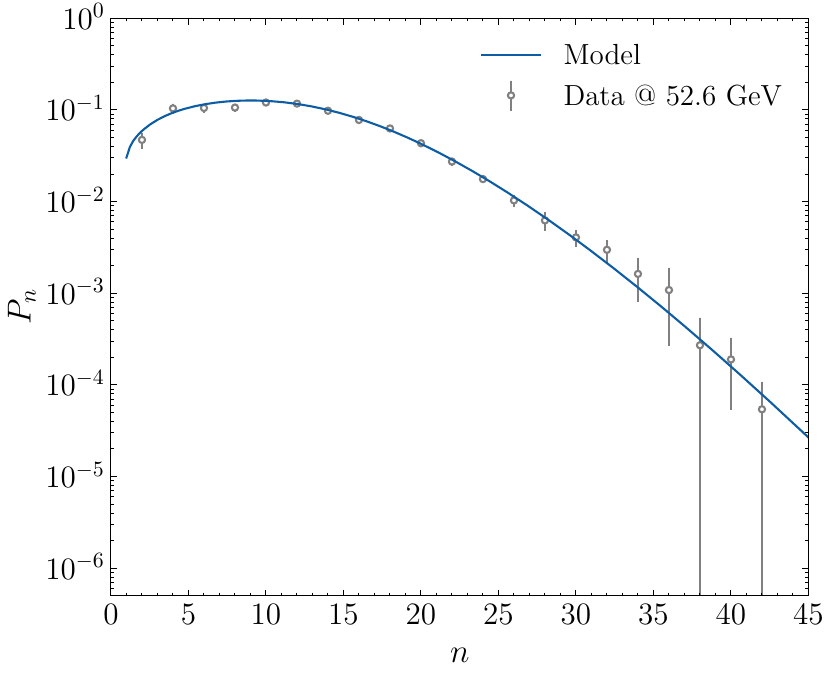}}
\subfloat{\includegraphics[height=0.4\textwidth]{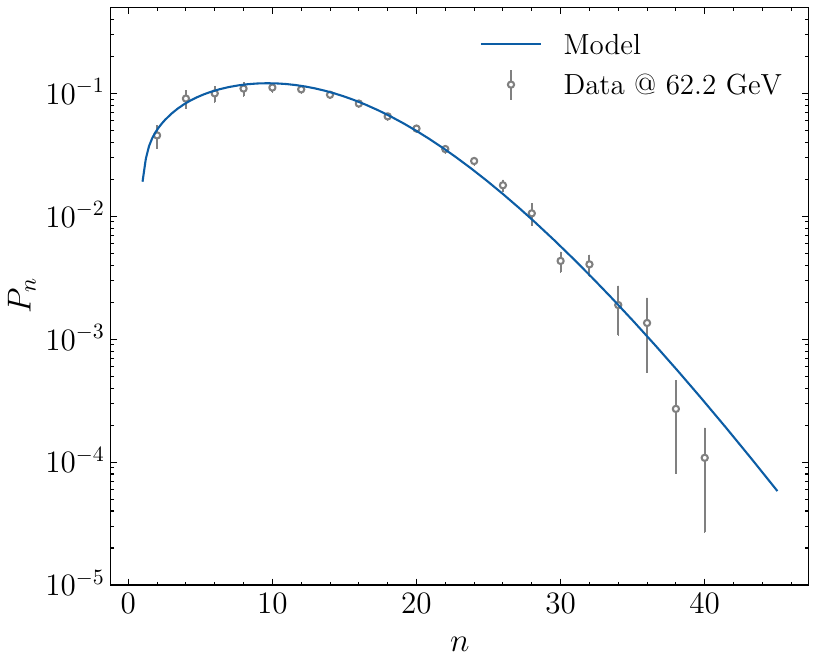}}
\caption{\label{fig:Pn_ISR} Multiplicity distributions for inelastic $pp$ data at $\sqrt{s}=30.4, 44.5, 52.6$ and $62.2$ GeV compared with theoretical expectations.}
\end{figure}

\begin{figure}[!htpb]
\centering
\subfloat{\includegraphics[height=0.4\textwidth]{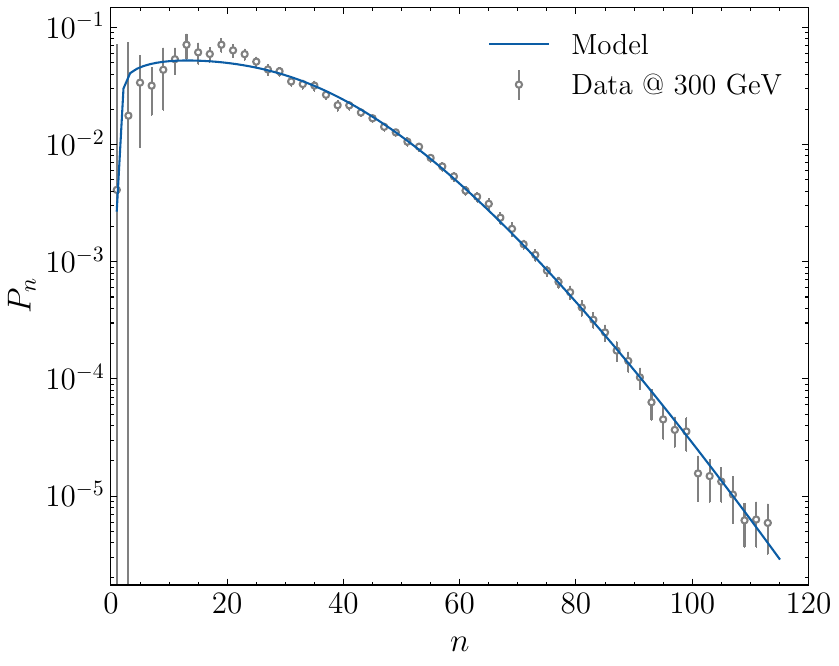}}
\subfloat{\includegraphics[height=0.4\textwidth]{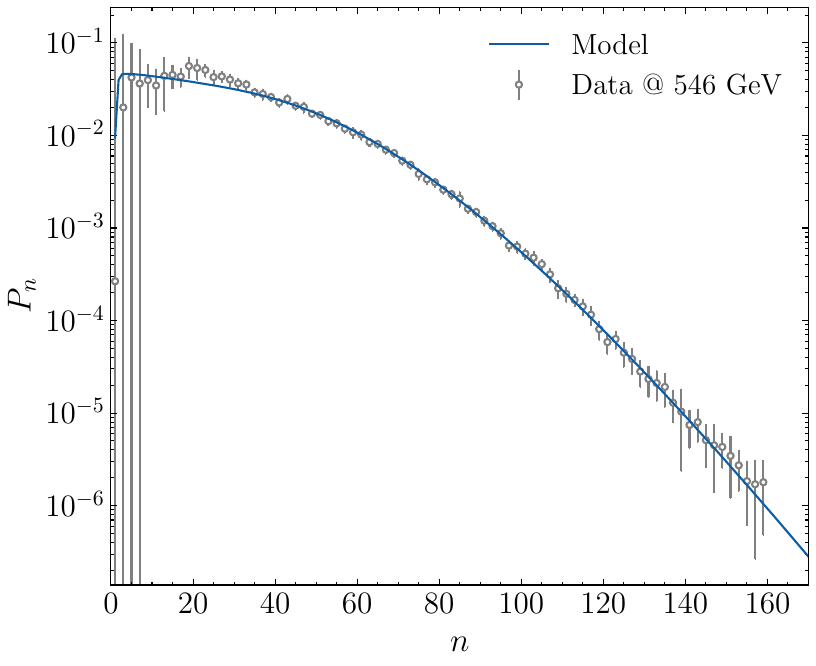}} \\
\subfloat{\includegraphics[height=0.4\textwidth]{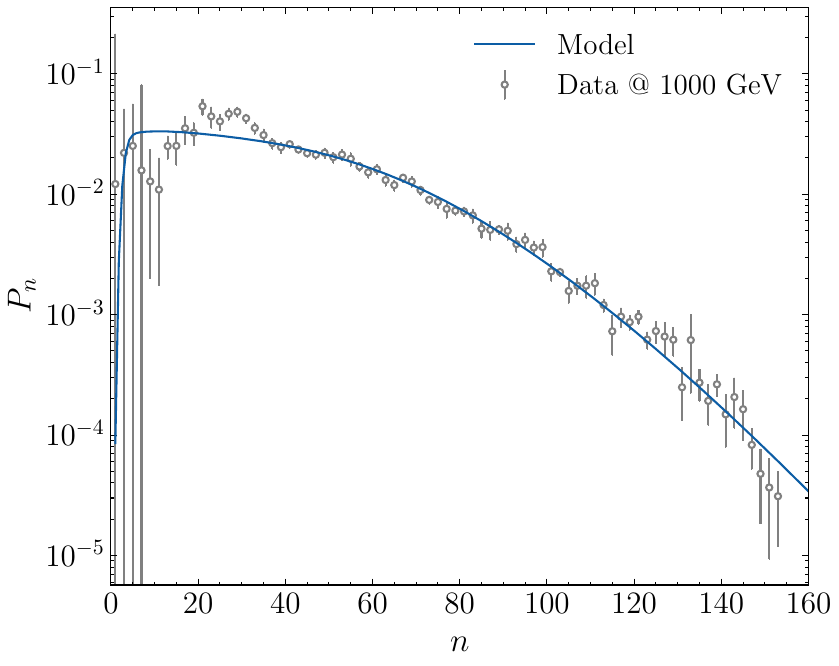}}
\subfloat{\includegraphics[height=0.4\textwidth]{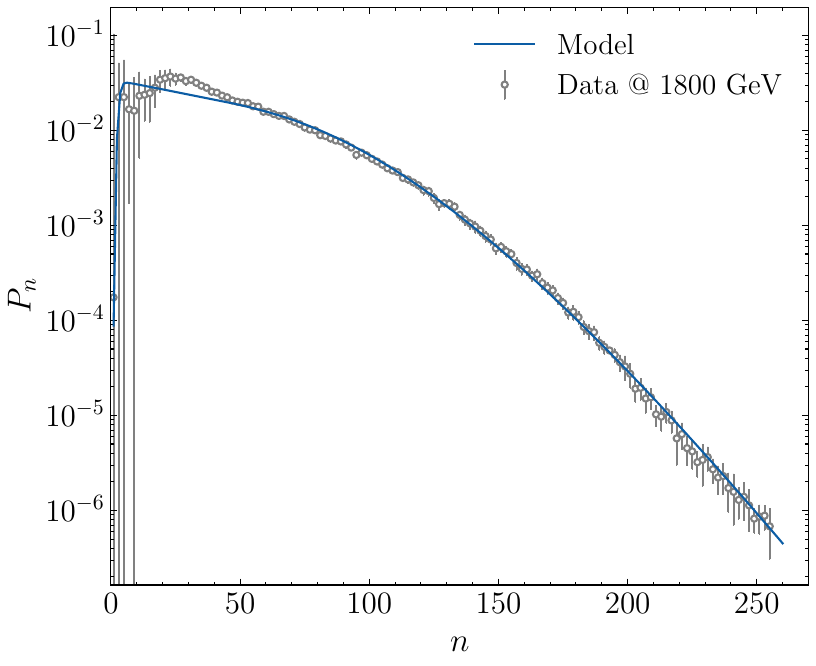}}
\caption{\label{fig:Pn_LHC} Multiplicity distributions for inelastic $\bar{p}p$ data at $\sqrt{s}=300, 546, 1000$ and $1800$ GeV compared with theoretical expectations.}
\end{figure}

Using the master formula (\ref{eq:master_f}), the outcomes of the fitting procedure of the multiplicity distributions data across a wide range of energies are provided in Fig.~\ref{fig:Pn_ISR}, Fig.~\ref{fig:Pn_LHC} and, Table~\ref{tab:bf_Pn}, respectively, where the values of the parameters $\lambda$ and $\langle n(s) \rangle$ obtained in each fitting procedure, along with the corresponding $\beta(s)$, are furnished, as well as the different $\chi^2/dof$ values. 

\begin{table}[!htpb]
\begin{center}
\resizebox{0.7\textwidth}{!}{%
\begin{tabular}{|c|c|c|c|c|}
\hline
$\sqrt{s}$ [GeV] & $\lambda$ & $\beta(s)$ & $\langle n(s) \rangle $ & $\chi^{2}/DOF$ \\
\hline
30.4 & 0.2837 $\pm$ 0.0133 & $1.6014 $ & 9.0583  $\pm$ 0.1452 & $1.2714$ \\
44.5 & 0.2720 $\pm$ 0.0119 & $ 1.6308$ & 10.5900 $\pm$ 0.1339 & $0.6170$ \\
52.6 & 0.2775 $\pm$ 0.0104 & $1.6430 $ & 11.2885 $\pm$ 0.1299 & $0.6443 $ \\
62.2 & 0.2709 $\pm$ 0.0107 & $1.6547 $ & 12.0090 $\pm$ 0.1551 & $1.3303 $ \\
300  & 0.3695 $\pm$ 0.0113 & $1.7327 $ & 23.4646 $\pm$ 0.2915 & $0.6046 $ \\
546  & 0.4618 $\pm$ 0.0148 & $1.7420 $ & 27.7185 $\pm$ 0.4145 & $0.3406 $ \\
1000 & 0.4230 $\pm$ 0.0110 & $1.7362 $ & 36.6360 $\pm$ 0.4042 & $1.5301 $ \\
1800 & 0.4836 $\pm$ 0.0070 & $1.7132 $ & 42.7217 $\pm$ 0.3184 & $1.2004 $ \\

7000 & 0.5936  & $1.5833 $ & $ 73.7117 $ & $ -  $ \\
14000 & 0.6597  & $ 1.4716 $ & $ 96.0705 $ & $ -  $ \\
\hline
\end{tabular}}

\end{center}
\caption{\label{tab:bf_Pn}Values of the $\lambda$ parameter and $\langle n(s) \rangle$ resulting from fits to the $P_{n}$ data. The values of $\beta(s)$ were obtained from eq. (\ref{eq:beta}).} 
\end{table}

According to Fig.~\ref{fig:Pn_ISR}, Fig.~\ref{fig:Pn_LHC}, and the different $\chi^2/dof$ values, our model gives a reasonable description of the different multiplicity distributions at each energy.
 Moreover, it allows to predict the multiplicity distribution $P_{n}$ at LHC energies by evaluating the energy dependence of the parameter $\lambda$ using an appropriate function $\lambda(s)$.

\begin{figure}[!htpb] 
\centering
\subfloat{\includegraphics[height=0.4\textwidth]{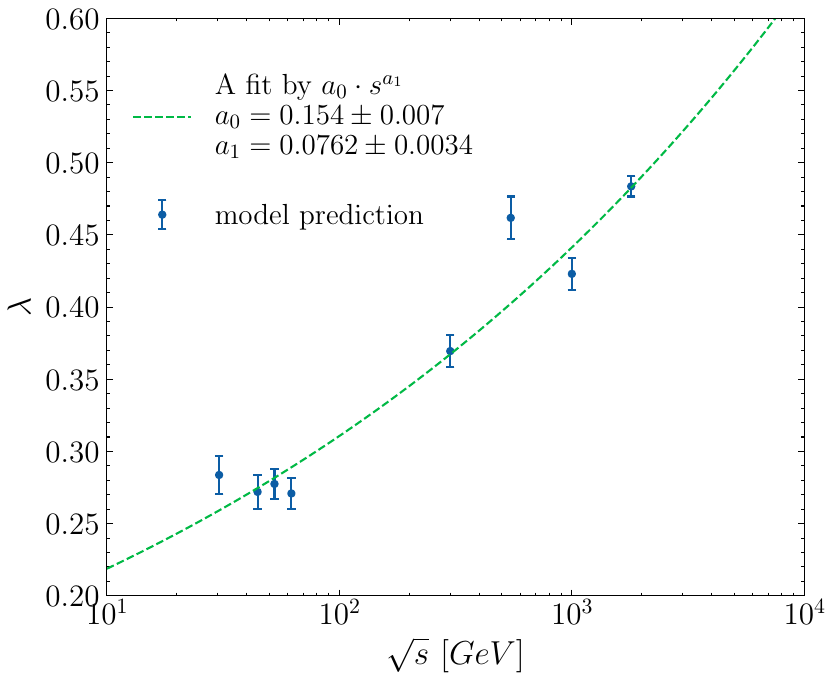}}
\subfloat{\includegraphics[height=0.4\textwidth]{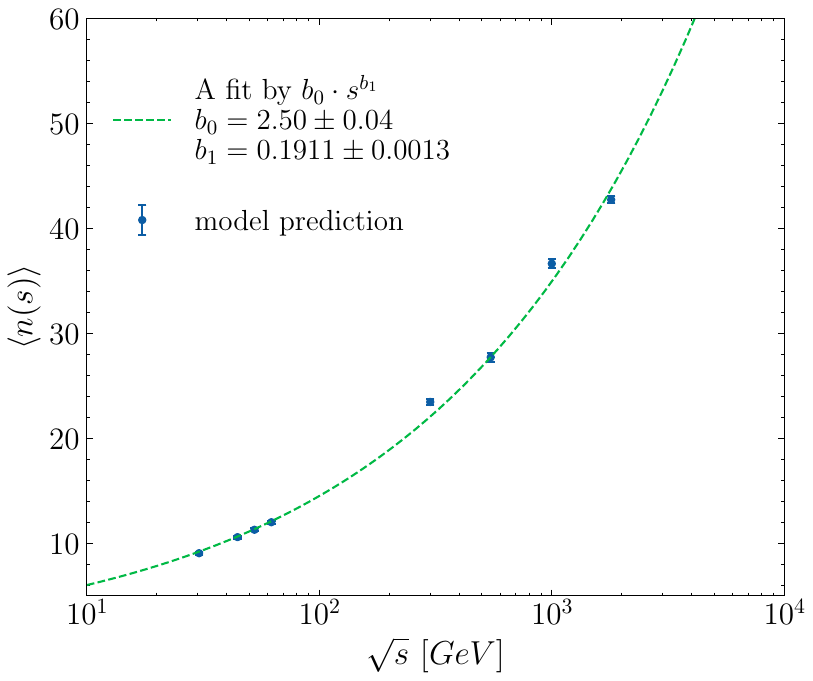}} 
\caption{\label{fig:lambda_mean_multip} The energy dependence of the parameter $\lambda$ (left panel) and the mean multiplicity (right panel)}
\end{figure}

As can be seen from Fig.\ref{fig:lambda_mean_multip} (left panel), the behaviour of this parameter is energy-dependent and rapidly increases with increasing energy. This energy dependence can be aptly described by the following function:
\begin{equation}
\lambda(s) = a_{0} \; s^{a_{1}}
\end{equation} where the values $a_{0} = 0.154 $, $ a_{1} = 0.0762 $ were determined by a careful $\chi^{2}$ analysis. Thus, based on this function, the $\lambda$ values at $7$ and $14$ TeV are retrieved and then displayed in Table \ref{tab:bf_Pn}.

Similarly, the estimates for the average multiplicity of hadrons at LHC energies can be derived. As depicted in Figure ~\ref{fig:lambda_mean_multip} (right panel), this parameter exhibits a rapid increase with the energy $s$, and its energy dependence can be consistently described using this function:

\begin{equation}
\langle n(s) \rangle= b_{0} \; s^{b_{1}},
\end{equation} where the values of $b_{0}$ and $b_{1}$ are $2.5$ and $0.1911$, respectively. These coefficients were determined through a rigorous $\chi^{2}$ analysis. Hence, by using this dependence, one can determine the values of $\langle n(s) \rangle$ at $7$ and $14$ TeV, as illustrated in Table \ref{tab:bf_Pn}.

An intriguing aspect of our findings lies in the remarkable accord between our result concerning the energy dependence of the hadron mean multiplicity and that obtained by Troshin and Tyurin with their model for multi-particle production with antishadowing \cite{Troshin_2003}, adding to the growing body of evidence that supports the fundamental principles underlying the $U$-Matrix approach. This alignment is highlighted by the following equation \cite{Troshin_2003} :

\begin{equation}
\langle n(s) \rangle = 2.328 \; s^{0.201}, 
\label{eq:TT_eq}
\end{equation}

This also implies that this approach is highly predictive and can accurately describe and interpret multi-particle production in different energy regimes. Besides, we should emphasize that the power-law energy dependence of the hadron mean multiplicity is often regarded as a prominent feature observed in different models and consistent with experimental data from heavy ion collisions \cite{Barshay:2001wp, Singh:2013fha} and this alignment in results further reinforces this assumption. 

Having tuned our model with all parameters obtained from the best fits, we can now rely on its potential extrapolations to novel collision energy regimes and investigate various phenomena, such as the KNO scaling violation and the correlation of final state particles, as will be presented in the subsequent sections.

\subsection{KNO scaling violation}

\begin{figure}[!htpb] 
\centering
\subfloat{\includegraphics[height=0.4\textwidth]{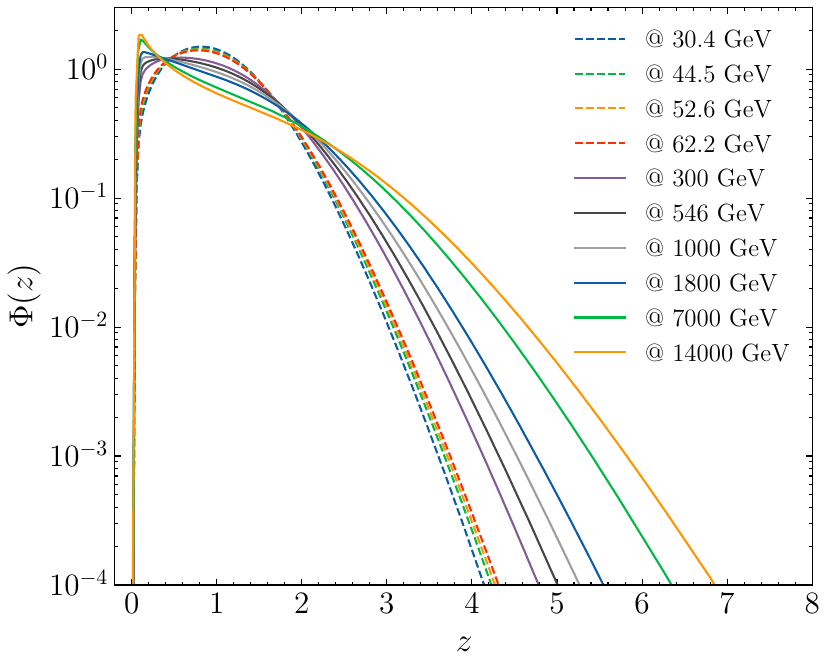}}
\subfloat{\includegraphics[height=0.4\textwidth]{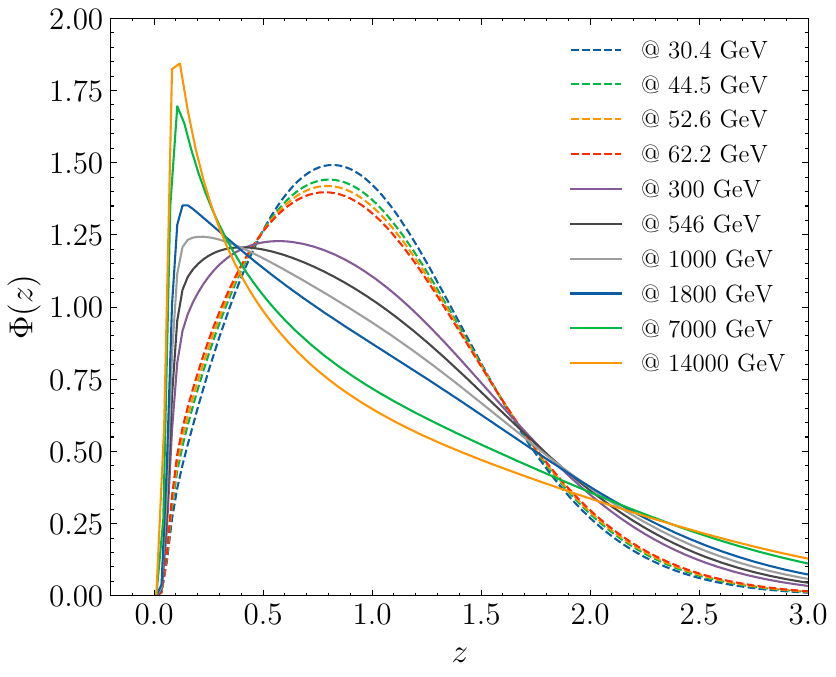}} 
\caption{\label{fig:kno_multip} The multiplicity distributions from ISR to LHC energies in full-phase space (left panel). In addition to the logarithmic view, the right panel shows a linear scale and a zoom into the low-multiplicity region.}
\end{figure}

Using our model, the KNO scaling violation was also examined. The predictions for the full-phase space multiplicity distribution in $p + p(\bar{p})$ collision, in KNO form at various energies, spanning from ISR to LHC ones, are displayed in Fig.~\ref{fig:kno_multip}. Aside from the logarithmic view in the left panel, a linear scale along with a zoom into the low-multiplicity region are provided in the right panel.

The left panel of Fig.~\ref{fig:kno_multip} shows that the high-multiplicity tail rises with increasing energy. At the same time, by looking at the right panel, we can see that the maximum of the distribution shifts towards smaller values of $z$.

This interpretation demonstrates the dynamic behaviour of the system, especially as the energy level rises, and further validates the violation of the KNO scaling. Interestingly, we can also see that beyond the ISR energy range, the width of the distribution gets larger with increasing energy, which underscores the strong violation of the KNO scaling. It is worth noting that this finding resonates with experimental observations \cite{Grosse-Oetringhaus_2010}.

Most importantly, based on the picture that KNO scaling violation is an extension of geometrical scaling violation, we can also claim that the strong violation of the former stems from the strong violation of the latter, emphasizing the interconnected nature of these phenomena within the $U$-matrix representation and stressing the latter’s pivotal role in describing collision geometry and the processes of multi-particle production in hadron collisions.

\begin{figure}[!htpb] 
\centering
\subfloat{\includegraphics[height=0.4\textwidth]{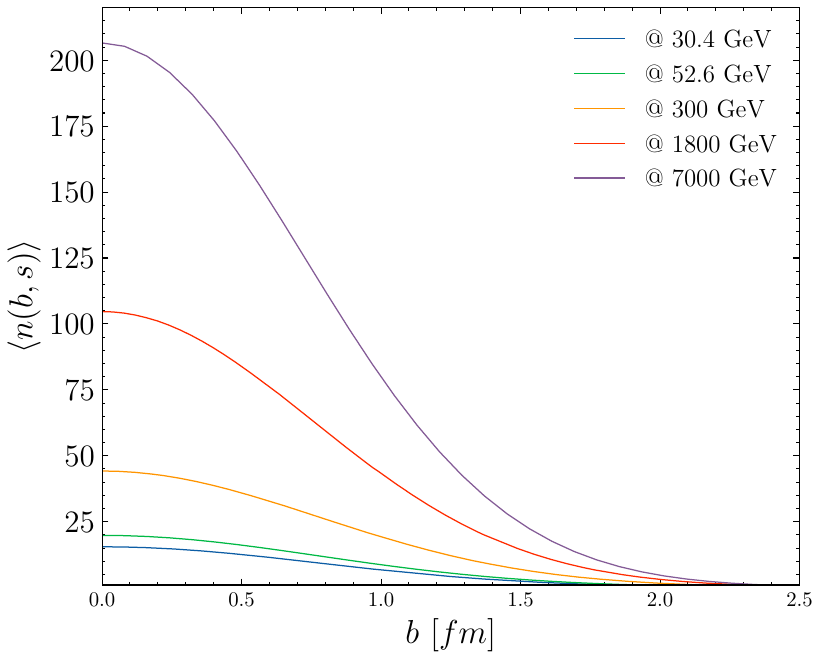}}
\subfloat{\includegraphics[height=0.4\textwidth]{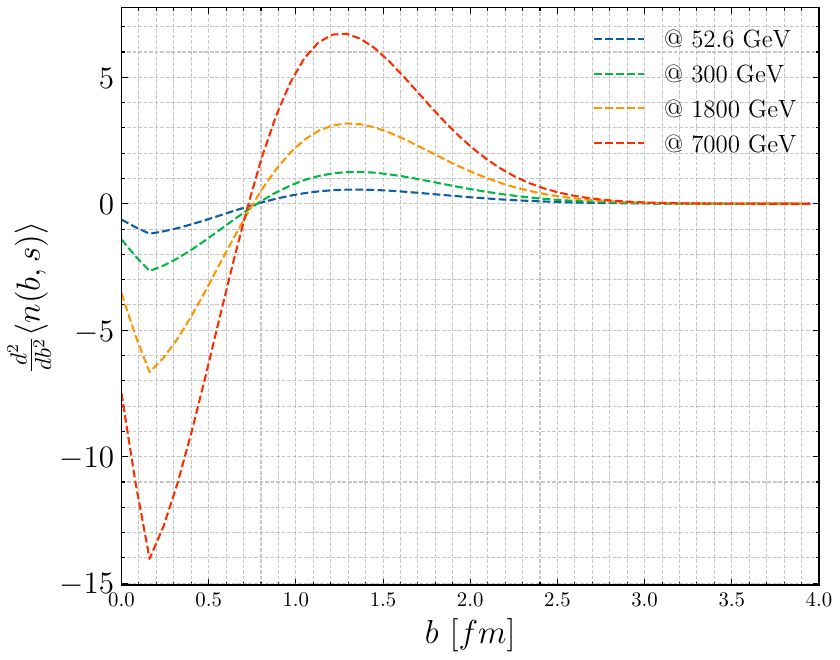}}
\caption{\label{fig:nbs} The energy dependence of the average multiplicity in the impact parameter space.}
\end{figure}

To further illustrate the role of the $U$-matrix scheme, we examined the average number of particles $\langle n(b, s)\rangle$ as a function of impact parameter $b$ for various collision energies, as its pattern offers insights into the collision geometry and the distribution of particles in the transverse plane. The result is illustrated in Fig.~\ref{fig:nbs} (left panel). Based on this figure, it is clear that, at central collision $(b = 0)$, the magnitude of the average number of particles increases with increasing energy. This is quite anticipated since central collisions yield more produced particles than peripheral collisions \cite{Yudong_1989}. This trend causes the tail of the multiplicity distribution to extend to higher values and eventually to a broader distribution as energy increases, indicating the possibility of rare high-multiplicity events.

A possible explanation for the broadening of the multiplicity distribution might be related to the effect of using the $U$-matrix scheme. Indeed, as previously illustrated in Fig.~\ref{fig:G_inel}, the disparity found between the eikonal and $U$-matrix schemes is noticeable at the central impact parameter, where the overlap of hadronic matter distributions is greater with the latter scheme. This discrepancy not only influences the inelastic overlap function, and hence the overall magnitude of multiplicity, but it also has a significant impact on the tail of the multiplicity distribution.

In addition, irrespective of the energy level, the average number of particles generally decreases as the impact parameter $b$ increases, which eventually leads to a decrease in the magnitude of the multiplicity distribution. This phenomenon aligns with the notion that larger impact parameters lead to less violent collisions, resulting in events with smaller multiplicities \cite{Yudong_1989}.

Furthermore, as Fig.~\ref{fig:nbs} manifests, the average number of particles is energy-dependent, increasing with higher collision energies at each impact parameter. This behaviour is also expected in high-energy physics experiments, where higher energies often result in increased particle production \cite{Grosse-Oetringhaus_2010}.

Fig.~\ref{fig:nbs} (left panel) shows that the curvature of the average number of particles changes, at a specific impact parameter, regardless of the energy level. More quantitatively, we show the second derivative of the average number of particles as a function of the impact parameter in Fig.~\ref{fig:nbs} (right panel), illustrating a change of sign of this function which demonstrates the existence of an inflection point at exactly $0.7$ fm. This change in the curvature could signify a shift in the particle production behaviour. For instance, at this specific impact parameter value, there might be a change in the interaction dynamics or the nature of the collision process. It should be noted that this result is in line with what has been reported in \cite{Beggio:2008zz}, where the inflection point was roughly estimated to be at around $1$fm. The fact that the inflection point is specifically at around $1$ fm suggests that there is something unique or significant about collisions occurring at this distance. This could be related to the characteristics of the colliding particles or the structure of the hadrons involved. This common behaviour underscores the notion of the critical phenomenon in hadronic interactions \cite{Iwao:1974fd}.

On the whole, the shape of the multiplicity distribution is influenced not only by the impact parameter and collision energy but also by the unitarisation scheme, particularly the $U$-matrix, which reinforces our claim that this distribution is scheme-dependent, as outlined in the introductory section. Besides, the tail of the multiplicity distribution gives insights into the rare but significant events that constitute the overall dynamics of high-energy collisions and hence highlights the importance of this scheme in multi-particle production processes.

In light of this result, the $U$-matrix scheme may prove to be a significant alternative for addressing multi-particle production challenges at ultra-high energies, including the muon puzzle in cosmic ray interactions at this energy level. In situations such as these, where extreme conditions and rare events are likely to play a significant role, scheme-dependent effects on multiplicity distribution become relevant.

Having said that, it is vital to consider this scheme in enhancing the existing hadronic interaction models to tackle several lingering issues in high and ultra-high energy physics. However, this is beyond the remit of this study.

\subsection{Hadronic multi-particle correlations}

Now that we have estimated the multiplicity distribution $P_{n}$ across various energies, it seems appropriate to characterize it in an attempt to obtain a better understanding of the dynamics of the particle production process in hadron collisions. In order to fulfil this purpose, the normalized ordinary higher-order moments $(q > 2)$ of this distribution are analyzed. Technically speaking, $P_{n}$’s moments of order $q$ are defined as follows:

\begin{equation}
C_q = M_q/M_1^q,    
\end{equation} and

\begin{equation}
M_q = \sum_{n=0}^{\infty} n^q P_n,
\end{equation}

\begin{figure}[!htpb] 
\centering
\subfloat{\includegraphics[height=0.6\textwidth]{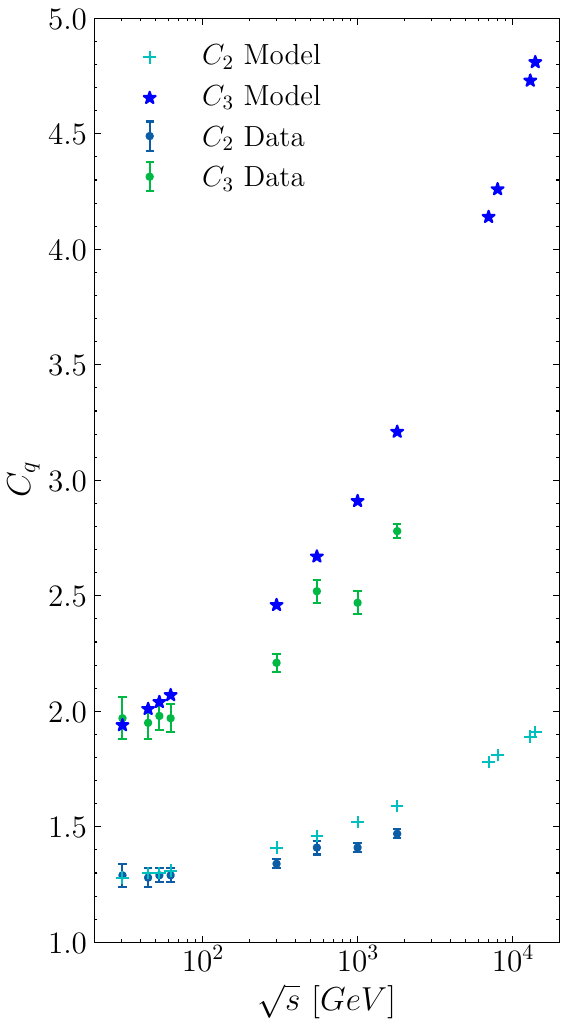}}
\subfloat{\includegraphics[height=0.6\textwidth]{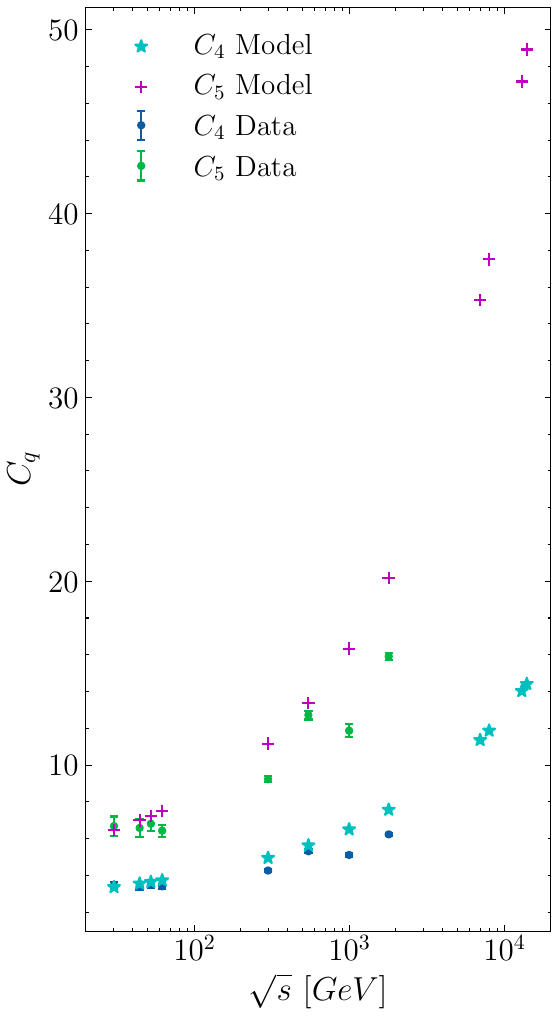}} 
\subfloat{\includegraphics[height=0.6\textwidth]{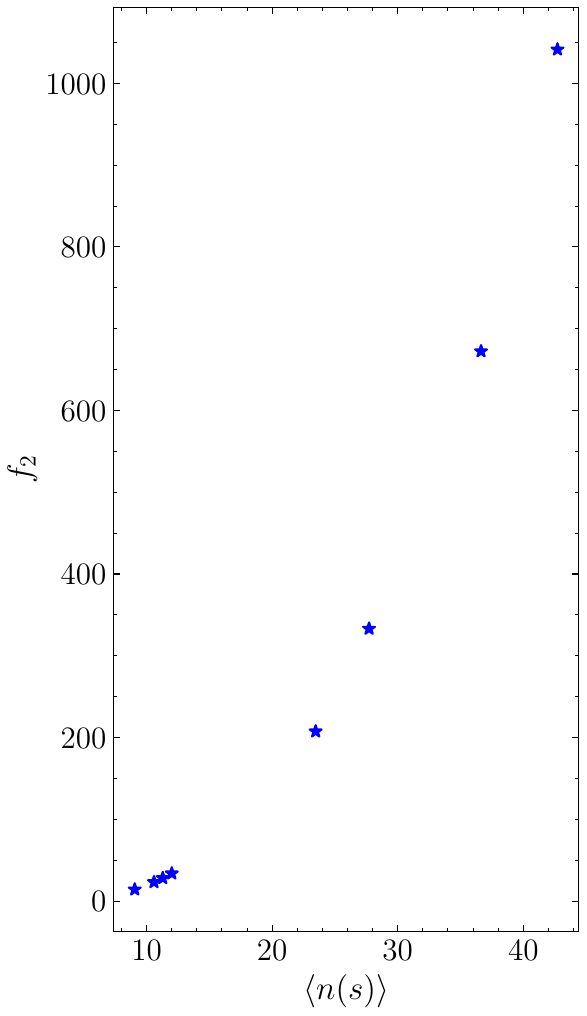}} 
\caption{\label{fig:C_q_f2} Experimental and Theoretical $C_{q}$ moments, q = 2, 3 (left panel), q = 4,5 (middle panel). $f_{2}$ moment versus the average number of produced particles (right panel).}
\end{figure}

Our results, along with their comparison with the experimental data \footnote{see compilation in \cite{Beggio:2017ran}}, are given in Table \ref{tab:Cq} and illustrated in Fig.~\ref{fig:C_q_f2} (left and middle panels).

\begin{table}[!ht]
\begin{center}
\resizebox{0.7\textwidth}{!}{%
\begin{tabular}{|c|c|c|c|c|}
\hline $\sqrt{s}$ $[GeV]$ & $C_{2}$ & $C_{3}$ & $C_{4}$ & $C_{5}$ \\
\hline
30.4 & 1.29 $\pm$ 0.05 & $1.97 \pm$ 0.09 & $3.45\pm 0.21$ & $ 6.68\pm 0.52$ \\
     & 1.28            & $  1.94 $          & $ 3.36        $ & $6.47 $         \\
\hline
44.5 & 1.28 $\pm$ 0.04 & $1.95 \pm$ 0.07 & $3.40\pm 0.17$ & $ 6.58\pm 0.47$ \\
     & 1.3            & $ 2.01 $          & $ 3.55        $ & $6.99 $         \\
\hline
52.6 & 1.29 $\pm$ 0.03 & $1.98 \pm$ 0.06 & $3.48\pm 0.15$ & $ 6.81\pm 0.42$ \\
     &  1.3           & $2.04 $          & $  3.64     $ & $7.24 $         \\
\hline
62.2 & 1.29 $\pm$ 0.03 & $1.97 \pm$ 0.06 & $3.40\pm 0.14$ & $ 6.43\pm 0.33$ \\
     &     1.31        & $ 2.07$          & $  3.73      $ & $7.51 $         \\
\hline
300  & 1.34 $\pm$ 0.02 & $2.21 \pm$ 0.04 & $4.26\pm 0.07$ & $ 9.23\pm 0.17$ \\
     &  1.41           & $ 2.46 $          & $   4.95    $ & $ 11.17$         \\
     \hline
546  & 1.41 $\pm$ 0.03 & $2.52 \pm$ 0.05 & $5.31\pm 0.10$ & $ 12.72\pm 0.24$ \\
     &  1.46           & $2.67 $          & $  5.63      $ & $ 13.38 $         \\
     \hline
1000 & 1.41 $\pm$ 0.02 & $2.47 \pm$ 0.05 & $5.11\pm 0.13$ & $ 11.87\pm 0.36$ \\
     &  1.52           & $2.91 $          & $  6.5       $ & $16.33 $         \\
     \hline
1800 & 1.47 $\pm$ 0.02 & $2.78 \pm$ 0.03 & $6.23\pm 0.07$ & $ 15.91\pm 0.21$ \\
     &  1.59         & $ 3.21 $          & $  7.57      $ & $ 20.18 $         \\
     \hline
7000 &  1.78           & $4.14 $          & $  11.36       $ & $35.3 $         \\
     \hline
8000 &   1.81          & $4.26 $          & $  11.87       $ & $37.51 $         \\
 \hline
13000 &   1.89          & $ 4.73$          & $   14.03     $ & $47.19 $         \\
 \hline
14000 &   1.91          & $4.81 $          & $   14.41     $ & $ 48.93 $         \\
\hline
\end{tabular}}

\end{center}
\caption{\label{tab:Cq} $C_{q}$ Moments: experimental data with error bar and theoretical predictions. Data points are from \cite{Beggio:2017ran}}  
\end{table}

Based on Fig.~\ref{fig:C_q_f2}, we can see that, as energy levels rise, our proposed model predicts a gradual increase in the ordinary higher-order moments, represented by $C_{2}$, $C_{3}$, $C_{4}$, and $C_{5}$. Surprisingly, while our predictions match with the data points within the ISR energy range, it is clear that the model overestimates the fluctuations and correlations in the multiplicity distribution with rising energy, notably above LHC energy. In order to further illustrate this overestimation, we computed the $f_{2}$ moment (or the two-particle correlation parameter), as a means of examining the correlation between pairs of particles during a collision event, which is defined by the following formula :

\begin{equation}
    f_{2} = < n (n - 1 )> - <n>^2 
\end{equation}

Interestingly, as illustrated in Fig.~\ref{fig:C_q_f2} (right panel), there is a noteworthy and sudden increase in the two-particle correlation parameter versus the average number of produced particles, indicating the existence of strong correlations among the charged particles. Consequently, we can infer that the model incorporates correlations in the final state, despite being constructed on the basis of independent particle production. So the question that arises is where this correlation emerges from. As the overall hadronic multiplicity distribution is constructed by summing contributions from parton–parton collisions occurring at each impact parameter weighted by the inelastic overlap function, this overestimation of correlation is linked to the weight in this superposition model, and hence to the unitarisation scheme, in comparison with the predictions provided by an eikonal geometrical independent string model \cite{Beggio:2017ran}.

Our model's outcomes of a pronounced KNO scaling violation, together with the unexpected overestimation of the fluctuations and correlations with increasing energy, can potentially be attributed to statistical fluctuations. Hence, we may claim that in this $U$-matrix representation, pomeron exchange may involve more intricate dynamics, such as collective effects, non-perturbative QCD dynamics, or other interactions, leading to different statistical fluctuations beyond a simple Poissonian eikonal summation \cite{Boreskov:2005ee}. 

\section{\label{sec:conclusions}Conclusions}

The first part of the results’ section was concerned with the description of the geometrical scaling violation. In fact, the energy evolution of the elastic-to-total cross-section ratio was investigated using both the eikonal and $U$-matrix schemes. The results have revealed that when energy rises, this ratio increases non-linearly and more rapidly with the $U$-matrix scheme than with the eikonal, implying a stronger violation of the geometrical scaling. This pronounced violation was understood in terms of the divergence in the behaviour of the inelastic overlap function, particularly at the central impact parameter, where the magnitude of this function is greater with the $U$-matrix scheme than the eikonal, regardless of the energy level.

The second part of the results’ section was devoted to the hadronic multiplicity distributions. Our model was tuned and all parameters were obtained from the best fits to various hadronic multiplicity distributions data over a wide range of energies. It has been found that the present model provides a reasonable description of the different multiplicity distributions at each energy. Interestingly, our findings about the energy dependence of the hadron mean multiplicity agree well with Troshin and Tyurin's analyses of the multiparticle production in the antishadowing model \cite{Troshin_2003}. This agreement adds to the growing body of evidence supporting the fundamental principles underlying the $U$-Matrix approach, which ensures reliable extrapolations to novel collision energy regimes.

Based on our model, the KNO scaling violation was investigated as well. Our results, related to the behaviour of the tail, as well as the maximum of the multiplicity distribution, and especially to the strong KNO scaling violation, are in line with the experimental findings \cite{Grosse-Oetringhaus_2010}. The broadening of the multiplicity distribution has been also confirmed with the behaviour of the particles’ distribution in the transverse plane at various collision energies. Besides, the interesting finding of an inflection point in the average number of particles’ curve irrespective of the energy level, corroborates with another finding (ref) and highlights the concept of a critical phenomenon in hadronic interactions.

Besides, the normalized ordinary higher-order moments $(q > 2)$ of the multiplicity distribution were analyzed. Another surprising result was related to our model’s overestimation of the fluctuations and correlations in the multiplicity distribution as energy rises, notably above LHC energy. It is argued that this overestimation, is linked to the weight in this superposition model, and hence to the unitarisation scheme. It should be noted that the shape of the hadron multiplicity distribution is influenced not only by the impact parameter and collision energy but also by the unitarisation scheme, particularly the $U$-matrix. 

On the whole, the results of this study, such as those related to the strong geometrical scaling violation and its resultant pronounced KNO scaling violation, coupled with the overestimation of the charged particles’ correlation, can potentially be attributed to statistical fluctuations inherent in the $U$-matrix scheme. Hence, we may argue that in this $U$-matrix representation, pomeron exchange may involve more intricate dynamics, leading to different statistical fluctuations beyond a simple Poissonian eikonal summation.

As any research, this study is not without limitations. For instance, the assumption that each created string has an equal probability of turning into a pair of charged hadrons is acceptable but only as a first approximation. To refine the present model for a more reliable description, we can consider the introduction of charged particle correlations in the final state by incorporating, for example, string fusion, overlapping processes, or other interactions leading to correlated string dynamics as evidenced by theoretical predictions \cite{Beggio:2007} and experimental observations \cite{CMS:2019fur}. Additionally, it is important to investigate the impact of the implicit different statistical fluctuations in pomeron exchanges within the $U$-matrix representation, along with the correlation and/or collective behaviour in the production of final state particles.

The study concludes by proposing the $U$-matrix scheme as a noteworthy alternative for tackling challenges related to multi-particle production in hadron collisions, especially in scenarios where extreme conditions and rare events play a significant role in high and ultra-high energy physics. It also prompts an inquiry into the fundamental nature of pomeron exchange within the $U$-matrix scheme in comparison to the eikonal, despite that both schemes verify the unitarity constraint principle.

\acknowledgments{RO would like to thank Jean-René Cudell for his invaluable comments. RO also appreciates the insightful discussions with Paulo Cesar Beggio and Emerson Luna. Special thanks go to the computational resource provided by Consortium des Équipements de Calcul Intensif (CÉCI), funded by the Fonds de la Recherche Scientifique de Belgique (F.R.S.-FNRS) where a part of the computational work was carried out.}

\bibliographystyle{JHEP}
\bibliography{refs}

\providecommand{\href}[2]{#2}\begingroup\raggedright\begin{thebibliography}{10}

\bibitem{VANHOVE1971347}
L.~{Van Hove}, \emph{Particle production in high energy hadron collisions},
  \href{https://doi.org/https://doi.org/10.1016/0370-1573(71)90013-5}{\emph{Physics
  Reports} {\bfseries 1} (1971) 347}.

\bibitem{Giacomelli:2009fr}
G.~Giacomelli, \emph{{Particle production in high energy collisions}},
  \href{https://arxiv.org/abs/0901.4924}{{\ttfamily 0901.4924}}.

\bibitem{Liu_2022}
F.-H.~Liu, K.K.~Olimov and A.~Beraudo, \emph{Editorial: Particle production and
  system evolution in collisions from gev to tev},
  \href{https://doi.org/10.3389/fphy.2022.1093225}{\emph{Frontiers in Physics}
  {\bfseries 10} (2022) }.

\bibitem{Koley:2022tqg}
H.~Koley, A.~Mondal, S.~Kar, S.~Mukherjee, J.~Mondal, A.~Deb et~al.,
  \emph{{Different degrees of complexity in multiparticle production at the LHC
  energies with the transition from soft to hard processes}},  in \emph{{65th
  DAE BRNS Symposium on nuclear physics}}, 5, 2022
  [\href{https://arxiv.org/abs/2205.04429}{{\ttfamily 2205.04429}}].

\bibitem{DREMIN2001301}
I.~Dremin and J.~Gary, \emph{Hadron multiplicities},
  \href{https://doi.org/https://doi.org/10.1016/S0370-1573(00)00117-4}{\emph{Physics
  Reports} {\bfseries 349} (2001) 301}.

\bibitem{alves2019open}
R.~Alves~Batista, J.~Biteau, M.~Bustamante, K.~Dolag, R.~Engel, K.~Fang et~al.,
  \emph{Open questions in cosmic-ray research at ultrahigh energies},
  {\emph{Frontiers in Astronomy and Space Sciences} {\bfseries 6} (2019) 23}.

\bibitem{Grosse-Oetringhaus_2010}
J.F.~Grosse-Oetringhaus and K.~Reygers, \emph{Charged-particle multiplicity in
  proton–proton collisions},
  \href{https://doi.org/10.1088/0954-3899/37/8/083001}{\emph{Journal of Physics
  G: Nuclear and Particle Physics} {\bfseries 37} (2010) 083001}.

\bibitem{CMS:2019fur}
{\scshape CMS} collaboration, \emph{{Bose-Einstein correlations of charged
  hadrons in proton-proton collisions at $\sqrt{s} =$ 13 TeV}},
  \href{https://doi.org/10.1007/JHEP03(2020)014}{\emph{JHEP} {\bfseries 03}
  (2020) 014} [\href{https://arxiv.org/abs/1910.08815}{{\ttfamily
  1910.08815}}].

\bibitem{Pandav:2022xxx}
A.~Pandav, D.~Mallick and B.~Mohanty, \emph{{Search for the QCD critical point
  in high energy nuclear collisions}},
  \href{https://doi.org/10.1016/j.ppnp.2022.103960}{\emph{Prog. Part. Nucl.
  Phys.} {\bfseries 125} (2022) 103960}
  [\href{https://arxiv.org/abs/2203.07817}{{\ttfamily 2203.07817}}].

\bibitem{Shih_1997}
C.C.~Shih and J.-y.~Zhang, \emph{Particle production at high energy. i.
  geometrical orientated simulation of hadron-hadron collisions below 250
  gev/$c$}, \href{https://doi.org/10.1103/PhysRevC.55.378}{\emph{Phys. Rev. C}
  {\bfseries 55} (1997) 378}.

\bibitem{Beggio:1999}
P.C.~Beggio, M.J.~Menon and P.~Valin, \emph{{Scaling violations: Connections
  between elastic and inelastic hadron scattering in a geometrical approach}},
  \href{https://doi.org/10.1103/PhysRevD.61.034015}{\emph{Phys. Rev. D}
  {\bfseries 61} (2000) 034015}
  [\href{https://arxiv.org/abs/hep-ph/9908389}{{\ttfamily hep-ph/9908389}}].

\bibitem{Beggio:2007}
P.C.~Beggio and Y.~Hama, \emph{{A new scheme for calculation of the
  multiplicity distributions in hadronic interactions}}, {\emph{Braz. J. Phys.}
  {\bfseries 37} (2007) 1164}.

\bibitem{BEGGIO_2014}
P.~Beggio and E.~Luna, \emph{Cross sections, multiplicity and moment
  distributions at the lhc},
  \href{https://doi.org/https://doi.org/10.1016/j.nuclphysa.2014.06.016}{\emph{Nuclear
  Physics A} {\bfseries 929} (2014) 230}.

\bibitem{Bhattacharya:2020lac}
A.~Bhattacharya, J.-R.~Cudell, R.~Oueslati and A.~Vanthieghem, \emph{{Proton
  inelastic cross section at ultrahigh energies}},
  \href{https://doi.org/10.1103/PhysRevD.103.L051502}{\emph{Phys. Rev. D}
  {\bfseries 103} (2021) L051502}
  [\href{https://arxiv.org/abs/2012.07970}{{\ttfamily 2012.07970}}].

\bibitem{Vanthieghem:2021akb}
A.~Vanthieghem, A.~Bhattacharya, R.~Oueslati and J.-R.~Cudell,
  \emph{{Unitarisation dependence of diffractive scattering in light of
  high-energy collider data}},
  \href{https://doi.org/10.1007/JHEP09(2021)005}{\emph{JHEP} {\bfseries 09}
  (2021) 005} [\href{https://arxiv.org/abs/2104.12923}{{\ttfamily
  2104.12923}}].

\bibitem{GayDucati:2007fbs}
M.B.~Gay~Ducati, M.M.~Machado and M.V.T.~Machado, \emph{{Investigating
  diffractive W production in hadron-hadron collisions at high energies}},
  \href{https://doi.org/10.1142/S0218301307008811}{\emph{Int. J. Mod. Phys. E}
  {\bfseries 16} (2007) 2956}
  [\href{https://arxiv.org/abs/0706.3468}{{\ttfamily 0706.3468}}].

\bibitem{BOZZO1984392}
M.~Bozzo, P.~Braccini, F.~Carbonara, R.~Castaldi, F.~Cervelli, G.~Chiefari
  et~al., \emph{Measurement of the proton-antiproton total and elastic cross
  sections at the cern sps collider},
  \href{https://doi.org/https://doi.org/10.1016/0370-2693(84)90139-4}{\emph{Physics
  Letters B} {\bfseries 147} (1984) 392}.

\bibitem{ALNER1984304}
G.~Alner, K.~Alpgård, R.~Ansorge, B.~Åsman, K.~Böckmann, C.~Booth et~al.,
  \emph{Scaling violation favouring high multiplicity events at 540 gev cms
  energy},
  \href{https://doi.org/https://doi.org/10.1016/0370-2693(84)91666-6}{\emph{Physics
  Letters B} {\bfseries 138} (1984) 304}.

\bibitem{KOBA1972317}
Z.~Koba, H.~Nielsen and P.~Olesen, \emph{Scaling of multiplicity distributions
  in high energy hadron collisions},
  \href{https://doi.org/https://doi.org/10.1016/0550-3213(72)90551-2}{\emph{Nuclear
  Physics B} {\bfseries 40} (1972) 317}.

\bibitem{Itabashi}
K.~Itabashi, \emph{{Koba-Nielsen-Olesen Scaling, Geometrical Scaling and
  Barshay-Yamaguchi Scaling}},
  \href{https://doi.org/10.1143/PTP.54.1168}{\emph{Progress of Theoretical
  Physics} {\bfseries 54} (1975) 1168}.

\bibitem{Finkelstein:1988cu}
J.~Finkelstein, \emph{{AN IMPACT PARAMETER MODEL FOR MULTIPLICITY
  DISTRIBUTIONS}}, \href{https://doi.org/10.1007/BF01412591}{\emph{Z. Phys. C}
  {\bfseries 41} (1988) 167}.

\bibitem{lam:1982}
C.S.~{Lam} and P.S.~{Yeung}, \emph{{Possible connection between KNO and
  geometrical scaling}},
  \href{https://doi.org/10.1016/0370-2693(82)90709-2}{\emph{Physics Letters B}
  {\bfseries 119} (1982) 445}.

\bibitem{FAGUNDES201248}
D.~Fagundes, E.~Luna, M.~Menon and A.~Natale, \emph{Aspects of a dynamical
  gluon mass approach to elastic hadron scattering at lhc},
  \href{https://doi.org/https://doi.org/10.1016/j.nuclphysa.2012.05.002}{\emph{Nuclear
  Physics A} {\bfseries 886} (2012) 48}.

\bibitem{Oueslati:2023tjt}
R.~Oueslati, \emph{{A multi-channel U-Matrix model of hadron interaction at
  high energy}}, \href{https://doi.org/10.1007/JHEP08(2023)087}{\emph{JHEP}
  {\bfseries 08} (2023) 087}
  [\href{https://arxiv.org/abs/2305.03424}{{\ttfamily 2305.03424}}].

\bibitem{CHOU1982301}
T.~Chou and C.N.~Yang, \emph{Remarks on multiplicity fluctuations and kno
  scaling in pp collider experiments},
  \href{https://doi.org/https://doi.org/10.1016/0370-2693(82)90348-3}{\emph{Physics
  Letters B} {\bfseries 116} (1982) 301}.

\bibitem{Barshay_PRL}
S.~Barshay, \emph{Geometric derivation of the diffractive multiplicity
  distribution}, \href{https://doi.org/10.1103/PhysRevLett.49.1609}{\emph{Phys.
  Rev. Lett.} {\bfseries 49} (1982) 1609}.

\bibitem{Kuang_1983}
C.~Kuang-chao, L.~Lian-sou and M.~Ta-chung, \emph{Koba-nielsen-olesen scaling
  and production mechanism in high-energy collisions},
  \href{https://doi.org/10.1103/PhysRevD.28.1080}{\emph{Phys. Rev. D}
  {\bfseries 28} (1983) 1080}.

\bibitem{Beggio:2020dry}
P.C.~Beggio and F.R.~Coriolano, \emph{{Energy dependence of the inelasticity in
  $pp/p{\overline{p}}$ collisions from experimental information on charged
  particle multiplicity distributions}},
  \href{https://doi.org/10.1140/epjc/s10052-020-7919-5}{\emph{Eur. Phys. J. C}
  {\bfseries 80} (2020) 437}
  [\href{https://arxiv.org/abs/2004.06839}{{\ttfamily 2004.06839}}].

\bibitem{ARTRU197493}
X.~Artru and G.~Mennessier, \emph{String model and multiproduction},
  \href{https://doi.org/https://doi.org/10.1016/0550-3213(74)90360-5}{\emph{Nuclear
  Physics B} {\bfseries 70} (1974) 93}.

\bibitem{Donnachie:2013xia}
A.~Donnachie and P.V.~Landshoff, \emph{{$pp$ and $\bar pp$ total cross sections
  and elastic scattering}},
  \href{https://doi.org/10.1016/j.physletb.2013.10.068}{\emph{Phys. Lett. B}
  {\bfseries 727} (2013) 500}
  [\href{https://arxiv.org/abs/1309.1292}{{\ttfamily 1309.1292}}].

\bibitem{Selyugin:2007jf}
O.V.~Selyugin, J.R.~Cudell and E.~Predazzi, \emph{{Analytic properties of
  different unitarization schemes}},
  \href{https://doi.org/10.1140/epjst/e2008-00773-0}{\emph{Eur. Phys. J. ST}
  {\bfseries 162} (2008) 37} [\href{https://arxiv.org/abs/0712.0621}{{\ttfamily
  0712.0621}}].

\bibitem{Cudell:2008yb}
J.R.~Cudell, E.~Predazzi and O.V.~Selyugin, \emph{{New analytic unitarisation
  schemes}}, \href{https://doi.org/10.1103/PhysRevD.79.034033}{\emph{Phys.
  Rev.} {\bfseries D79} (2009) 034033}
  [\href{https://arxiv.org/abs/0812.0735}{{\ttfamily 0812.0735}}].

\bibitem{Drescher:2000ha}
H.J.~Drescher, M.~Hladik, S.~Ostapchenko, T.~Pierog and K.~Werner,
  \emph{{Parton based Gribov-Regge theory}},
  \href{https://doi.org/10.1016/S0370-1573(00)00122-8}{\emph{Phys. Rept.}
  {\bfseries 350} (2001) 93}
  [\href{https://arxiv.org/abs/hep-ph/0007198}{{\ttfamily hep-ph/0007198}}].

\bibitem{Breakstone_1984}
{\scshape Ames-Bologna-CERN-Dortmund-Heidelberg-Warsaw Collaboration}
  collaboration\href{https://doi.org/10.1103/PhysRevD.30.528}{\emph{Phys. Rev.
  D} {\bfseries 30} (1984) 528}.

\bibitem{ALEXOPOULOS1998453}
T.~Alexopoulos, E.~Anderson, N.~Biswas, A.~Bujak, D.~Carmony, A.~Erwin et~al.,
  \emph{The role of double parton collisions in soft hadron interactions},
  \href{https://doi.org/https://doi.org/10.1016/S0370-2693(98)00921-6}{\emph{Physics
  Letters B} {\bfseries 435} (1998) 453}.

\bibitem{Hatlo_2005}
M.~Hatlo, F.~James, P.~Mato, L.~Moneta, M.~Winkler and A.~Zsenei,
  \emph{{Developments of mathematical software libraries for the LHC
  experiments}}, \href{https://doi.org/10.1109/TNS.2005.860152}{\emph{IEEE
  Trans. Nucl. Sci.} {\bfseries 52} (2005) 2818}.

\bibitem{Troshin_2003}
S.M.~Troshin and N.E.~Tyurin, \emph{Multiparticle production in the model with
  antishadowing},
  \href{https://doi.org/10.1088/0954-3899/29/6/309}{\emph{Journal of Physics G:
  Nuclear and Particle Physics} {\bfseries 29} (2003) 1061}.

\bibitem{Barshay:2001wp}
S.~Barshay and G.~Kreyerhoff, \emph{{Related power law growth of particle
  multiplicities near mid-rapidity in central Au + Au collisions and in anti
  $\bar p(p)-p$ collisions}},
  \href{https://doi.org/10.1016/S0375-9474(01)01258-1}{\emph{Nucl. Phys. A}
  {\bfseries 697} (2002) 563}
  [\href{https://arxiv.org/abs/hep-ph/0104303}{{\ttfamily hep-ph/0104303}}].

\bibitem{Singh:2013fha}
R.~Singh, L.~Kumar, P.K.~Netrakanti and B.~Mohanty, \emph{{Selected
  Experimental Results from Heavy Ion Collisions at LHC}},
  \href{https://doi.org/10.1155/2013/761474}{\emph{Adv. High Energy Phys.}
  {\bfseries 2013} (2013) 761474}
  [\href{https://arxiv.org/abs/1304.2969}{{\ttfamily 1304.2969}}].

\bibitem{Yudong_1989}
H.~Yudong, W.~Guangjun, M.~Jun and T.~An, \emph{On the geometrical picture of
  multiparticle production in high-energy hadron-hadron collisions},
  \href{https://doi.org/10.1209/0295-5075/9/7/006}{\emph{Europhysics Letters}
  {\bfseries 9} (1989) 645}.

\bibitem{Beggio:2008zz}
P.C.~Beggio, \emph{{A multiparton model for p p / p anti-p inelastic
  scattering}},
  \href{https://doi.org/10.1590/S0103-97332008000500012}{\emph{Braz. J. Phys.}
  {\bfseries 38} (2008) 598}.

\bibitem{Iwao:1974fd}
S.~Iwao and M.~Shako, \emph{{Hadron-Hadron Interactions and their
  Interpretation as Critical Phenomena}}, .

\bibitem{Beggio:2017ran}
P.C.~Beggio, \emph{{Inelastic cross sections, overlap functions and ${C}_{q}$
  moments from ISR to LHC energies in proton interactions}},
  \href{https://doi.org/10.1088/1361-6471/aa51f5}{\emph{J. Phys. G} {\bfseries
  44} (2017) 025002} [\href{https://arxiv.org/abs/1701.08574}{{\ttfamily
  1701.08574}}].

\bibitem{Boreskov:2005ee}
K.G.~Boreskov, A.B.~Kaidalov, V.A.~Khoze, A.D.~Martin and M.G.~Ryskin,
  \emph{{The Partonic interpretation of reggeon theory models}},
  \href{https://doi.org/10.1140/epjc/s2005-02376-8}{\emph{Eur. Phys. J. C}
  {\bfseries 44} (2005) 523}
  [\href{https://arxiv.org/abs/hep-ph/0506211}{{\ttfamily hep-ph/0506211}}].

\end{thebibliography}\endgroup
\end{document}